# Chandra Multi-wavelength Project (ChaMP).
# I. First X-ray Source Catalog

D.-W. Kim, R. A. Cameron, J. J. Drake, N. R. Evans, P. Freeman, T. J. Gaetz, H. Ghosh,
P. J. Green, F. R. Harnden, Jr., M. Karovska, V. Kashyap, P. W. Maksym, P. W. Ratzlaff,
E. M., Schlegel, J. D. Silverman, H. D. Tananbaum, A. A. Vikhlinin, B. J. Wilkes,

(Smithsonian Astrophysical Observatory, Cambridge, MA 02138)

and

J. P. Grimes

(Center for Astrophysical Sciences, Johns Hopkins University, Baltimore, MD 21218-2686)

August 26, 2003

Abstract

The Chandra Multi-wavelength Project (ChaMP) is a wide-area (~14 deg$^2$) survey of serendipitous Chandra X-ray sources, aiming to establish fair statistical samples covering a wide range of characteristics (such as absorbed AGNs, high z clusters of galaxies) at flux levels ($f_X \sim 10^{-15} - 10^{-14}$ erg sec$^{-1}$ cm$^{-2}$) intermediate between the Chandra Deep surveys and previous missions. We present the first ChaMP catalog, which consists of 991 near on-axis, bright X-ray sources obtained from the initial sample of 62 observations. The data have been uniformly reduced and analyzed with techniques specifically developed for the ChaMP and then validated by visual examination. To assess source reliability and positional uncertainty, we perform a series of simulations and also use Chandra data to complement the simulation study. The false source detection rate is found to be as good as or better than expected for a given limiting threshold. On the other hand, the chance of missing a real source is rather complex, depending on the source counts, off-axis distance (or PSF), and background rate. The positional error (95% confidence level) is usually < 1″ for a bright source, regardless of its off-axis distance while it can be as large as 4″ for a weak source (~20 counts) at a large off-axis distance ($D_{\text{off-axis}}$ > 8′). We have also developed new methods to find spatially extended or temporary variable sources and those sources are listed in the catalog.

1. Introduction

The launch of the Chandra X-ray Observatory has opened a new era in X-ray astronomy. With its unprecedented, sub-arcsec spatial resolution (van Speybroeck 1997), in conjunction with its high sensitivity and low background, Chandra is providing new views of the X-ray sky 10-100 times deeper than previously possible (Weisskopf et al. 2000). Indeed, the cosmic X-ray background, whose populations have long been debated because the necessary spatial resolution was lacking, is now almost (~80%) resolved into discrete sources in deep Chandra observations, e.g., the CDF-N (Chandra Deep Field-North; Brandt et al. 2001), the CDF-S (Giacconi et al. 2001). Moretti et al. (2003) has recently reported an even higher fraction (~90%). However, the nature of these sources is still somewhat unclear (e.g., Hasinger et al. 1998). An absorbed AGN population is predicted by population synthesis models (e.g. Comastri et al. 1995, Gilli et al. 1999) as the Cosmic X-ray Background is much harder (a photon index of ~1.4) than typical AGNs which have a photon index of ~1.7 (e.g. Marshall et al. 1980, Fabian & Barcons 1992). There is some observational evidence supporting the existence of red, absorbed quasars (e.g., Kim & Elvis 1999; Wilkes et al. 2002; White et al. 2003). However, the hard sources in the deep surveys appear to be a mix of various types of narrow and broad line AGNs and apparently normal galaxies with very few of the expected type 2 AGN seen. The statistical importance of these various source types requires a large sample resulting from a wider area survey such as ChaMP. Additionally, with two highly successful X-ray observatories currently in orbit (Chandra and XMM-Newton), we will soon be able to address fundamental questions such as: whether the density and luminosity of quasars are evolving in time (e.g., Miyaji et al. 2000, Cowie et al. 2003), and how clusters of galaxies form and evolve (e.g., Rosati, et al. 2002a). We will also discover whether rare, but important objects have been missing from previous studies (e.g., blank field sources discussed in Cagnoni et al. 2002).

To take full advantage of the rich dataset available in the Chandra public archive, we have initiated a serendipitous X-ray source survey, the Chandra Multi-wavelength Project (ChaMP). Owing to the high spatial resolution, identification of X-ray sources is far less ambiguous than in previous missions where many counterparts were often found within typical error circles (at least ~10 times larger). Additional information and artificial selection criteria are no longer required, leaving little bias. The ChaMP, although not as deep as the CDF, covers a wide area (~ 14 deg$^2$) and can provide an order of magnitude more sources at intermediate flux levels ($F_x \sim 10^{-14} – 10^{-15}$ erg sec$^{-1}$ cm$^{-2}$) than either the Chandra deep surveys or the previous missions (see Figure 1 and section 2). An additional advantage of a wide area survey is the ability to investigate field-to-field variations of the number density of cosmic (background) sources, which may trace filaments and voids in the underlying large-scale structure, or if not detected, constrain the hierarchical structure formation.

In this paper (paper I), we describe our data reduction and analysis methods uniquely developed for this project and present the first catalog obtained with an initial sample of 62 Chandra observations. In paper II (Kim et al. 2003), we present the results of X-ray source properties by producing the Log(N)-Log(S) relation and X-ray colors, and by comparing with data at other wavelengths. In an accompanying paper (Green et al. 2003), we present the first results of deep optical follow-up observations.

This paper is organized as follows. In section 2, we describe how Chandra observations were selected for the ChaMP study. We present in detail the data reduction and analysis techniques in section 3. We have complemented our analysis with an extensive set of simulations in order to quantitatively assess the detection probability (section 4) and positional accuracy (section 5), which are both critical to firmly establish statistical properties of X-ray sources. In section 6, we present the first ChaMP X-ray source catalog and describe its contents.

2. ChaMP Field Selection

We have carefully selected Chandra fields which are best-suited for ChaMP science. Our selection criteria are often distinct from the purpose for which the observations were originally intended. An ideal field is one for which the PI is interested in a small object in the center of the field, providing the largest sky area available for a sensitive survey. The ChaMP would then assemble sources from the large unstudied area outside the main target to be processed and analyzed uniformly. Optical images and spectra are then added to create a wide field multi-wavelength survey. A number of criteria were used in constructing the survey area:

- Only fields more than 20° from the Galactic plane were selected in order to keep the Galactic extinction low. The latitude limit corresponds to $N(H) < 6 \times 10^{20}$ cm$^{-2}$, or $E(B-V) < 0.1$ mag.
- Only ACIS imaging (i.e., no grating) fields were used; HRC images were omitted. Fields with instrumental complications (eg., a bias map missing in telemetry or corrupted, energy filtering, or spatial windows) were omitted, as were fields using sub-arrays or continuous clocking.
- Fields dominated by large extended sources (covering more than 10% of field of view) in either the optical or X-ray were omitted because these would mask faint sources. The extended source area is determined by the contour where the signal falls to 2 times the background (in case of ROSAT images) or to 10-sigma above the background (in case of POSS). RASS data were used where ROSAT pointed images were not available. Supernova remnants were automatically omitted. Additionally, to insure that the field of view would not be heavily contaminated by galaxies belonging to a cluster, we also exclude fields containing clusters with $z < 0.3$ (1 Mpc at z=0.3 corresponds to 3 arcmin.)
- Planetary observations were omitted.
- Observations closer than 10.8° from the center of the LMC, 5.3° from the SMC, and 3.2° from M31 were omitted.
- Fields intended by the PI for surveys were omitted. These include ELAIS:N1.1 and ELAIS N1.2, the Chandra Deep Field North and South, Lockman hole and MBM12.

Based on these selection criteria, 137 fields were selected among Chandra AO1 and AO2 observing periods. Of these, 62 ACIS observations, which are available to us either through the public archive or by PI's pre-approval, have been processed and presented in this paper. They consist of 21 ACIS-I and 41 ACIS-S observations based on whether the aim point falls on I3 (or CCDID=3) or S3 (or CCDID=7), respectively (see Figure 6.1 in Chandra Proposer's Observatory Guide, POG). In Table 1, we list the basic observational parameters of the selected Chandra observations including observation dates, the aim points (ra, dec), CCD chips used and effective exposure times. The effective exposure time was estimated after removing background flares within the main chip where the aim point lies (see Section 3.1).

We estimate the number of sources likely to be detected in the full ChaMP by assuming the combined 0.5-2~keV Log(N)-Log(S) from the CDF-South 1Msec ACIS-I and ROSAT/PSPC Lockman Hole 207ksec results (Rosati et al. 2002b and Hasinger et al. 1998, respectively) and applying this to each Chandra image in our full survey (see Green et al. 2002). We simulate the number of sources that would be detected in a given Chandra image by randomly populating the CCDs with sources derived from the Log(N)-Log(S), assuming an average spectral slope $\Gamma_{ph}= 1.4$. When the full ChaMP sample is complete, we thereby predict ~6000 X-ray detections. The number of predicted detections as a function of soft X-ray flux is shown in Figure 1 for the full ChaMP sample, the deep ROSAT sample (Lockman Hole; Hasinger et al. 1998)) and the combined CDFs (2 Msec North and 1 Msec South). Also shown in Figure 1 is the estimated sky area sensitive to a given flux limit. It is clear that the full ChaMP provides unprecedented coverage in the intermediate flux regime.

3. X-ray data analysis

3.1. Data Correction and Data Screening

We have developed a ChaMP-specific pipeline (called XPIPE) to uniformly reduce the Chandra data and to generate homogeneous data products. XPIPE was built mainly with CIAO (v2.3) tools (http://cxc.harvard.edu/ciao). An example of XPIPE products is seen in Figure 2. Level 2 data products generated by the Chandra X-ray Center (CXC) standard data processing are taken as an input and further processed by XPIPE for additional data correction and data screening. As CXC has reprocessed old data observed early in the mission with more reliable software and calibration data (http://cxc.harvard.edu/udocs/reprocessing.htm), we use Rev. 2 (or higher) data products whenever available. Standard data screening applied in the CXC pipeline processing excludes events with ASCA grade = 1, 5 and 7 (which are mostly cosmic ray events) and events with status bits set to non-zero. Screening by the status bit excludes bad pixels and columns among many other instrumental effects. A full description of ACIS status bit can be found in http://cxc.harvard.edu. In some cases, however, bad pixels and columns that are not listed in the calibration database (hence not excluded in CXC processing) are still visible, particularly in the data taken at an ACIS temperature of –110°C or –100°C (observed earlier than Jan. 2000; marked in Table 1) and in the S4 chip (CCDID=8). An example of bad columns is shown in Figure 3-a where a series of false sources along a CCD column is clearly seen near the bottom of the figure. We checked each image by plotting in chip coordinates for which the bad column is not blurred by the dither (see Section 3.5). In some observations, a finite number of bias values are corrupted and they are slightly lower for a given node in a given CCD. Since they are repeatedly subtracted in each exposure frame, the pixels with corrupted bias values appear to be hot pixels de-dithered with an aspect Lissajous pattern. In this case, several sources are spuriously detected at the corners of the Lissajous pattern of a single bad pixel (see Figure 3-b). They are identified by eye and those false sources are flagged in the ChaMP database (see Section 3.5). Some CXC data have been processed with outdated calibration data, such as the ACIS gain file used to produce PI/energy columns and the alignment file used to determine absolute positions. We re-apply new, updated calibration files whenever necessary.

Cosmic rays or charged particles sometimes leave residual charges in CCD pixels and they are repeatedly identified as a valid event in the same CCD pixels. This effect, known as afterglow, was first noticed by discovering very sharp point-like sources at large off-axis angles. To correct this, CXC has added a new tool to the pipeline processing (August 2000 in CXCDS release R4CU5UPD8). The CXC and the ACIS Instrument team (http://www.astro.psu.edu/xray/acis) note a side effect of this correction – the algorithm may remove real events up to a few percent, particularly for a bright source. Since this side effect mostly affects very bright sources, it will not alter results of source detection. In the ChaMP, we apply this correction so that any false source should be excluded. As the investigation of the effect of afterglow continues, if it is necessary to counter-correct events from bright sources, we will do so.

It is also known that the charge cascades (also called 'blooms') caused by interacting with cosmic rays effectively reduce the detector efficiency by as much as a few % (see http:// asc.harvard.edu/ ciao/ caveats). As the effects of the blooms and the workaround are being investigated, we will correct this effect, if necessary, when it is better understood.

The S4 chip (or CCDID=8) is known to suffer from an instrumental effect called streaking. This may be corrected by applying a de-streaking tool provided by John Houck (also available in the CIAO package). Figure 4 illustrates the pronounced difference before and after the de-streaking correction. While the correction works in most cases, during our visual inspection (Section 3.5), we recognized that streaking is not fully corrected in every case and that bad columns perhaps mixed with the streaking problem appear in a few observations. Therefore, we do not use the sources detected in the S4 chip for further analyses in this paper.

The ACIS background is known to vary significantly (http://cxc.harvard.edu/cal/Acis/Cal_prods/bkgrnd/current/index.html). The count rate can increase by a factor of up to 100 within a single observation. Background flares are more prominent in back-illuminated chips (S3 and S1) than front-illuminated chips. Typical examples of background light curves for a BI and an FI chip are shown in Figure 5. The origin of background flares is not known (e.g. Grant et al. 2001); low energy ($< 100$ keV) protons may be responsible (Struder et al. 2001). Because the source detection probability strongly depends on the background rate (section 4), we do not use the data obtained during background flares. After making a background light curve, we exclude those time intervals beyond 3-sigma fluctuation above the mean background count rate. The mean rate is determined iteratively after excluding the high background intervals. Given different characteristics between BI and FI chips, it is applied per ACIS CCD. The GTI (Good Time Interval) extension table is then updated accordingly so that the same CIAO tools can be applied with and without data screening by the background rate. Figure 6 shows a histogram of effective exposure times for BI and FI chips after data screening by the high background rate for the 62 observations reported here. While the loss is minimal for FI chips, it can be as large as 50% for BI chips. Among the 62 observations, the average effective exposure time was reduced by 18% in BI and 5% in FI chips.

3.2. Source Detection - **wavdetect**

To detect X-ray sources, we apply a wavelet detection algorithm, called **wavdetect**, available in the CIAO software package (Freeman et al. 2002). Because **wavdetect** is more reliable in finding individual sources in crowded fields and in identifying extended sources than the traditional celldetect algorithm, we have selected **wavdetect** as the main detection tool in the ChaMP. We run **wavdetect** repeatedly in three energy bands (B, S and H – see table 2) to quantify detections and upper limits in each sub energy band. After performing various tests to find the most efficient parameters used with **wavdetect**, we select a significance threshold parameter of $10^{-6}$ which corresponds to 1 possibly spurious pixel in one CCD (see section 4.1 for more discussion of this type I error) and a scale parameter of 7 steps between 1 and 64 pixels (1 pixel = 0.492 arcsec) to cover a wide range of source sizes, accommodating extended sources and the variation of the PSF as a function of off-axis distance ($D_{off-axis}$). For the remaining parameters, we used the default values given in CIAO (see section 4 for more discussion on **wavdetect** performance.) To avoid finding spurious sources, most often located at the edge of the field of view, we used an exposure map generated for each CCD at an energy of 1.5 keV with an appropriate aspect histogram (see http://cxc.harvard.edu/ciao/threads/ expmap_acis_single) and required a minimum of 10% of the on-axis exposure.

As shown in Figure 2, **wavdetect** performs well in identifying both point sources and extended sources (e.g., an extended source toward the lower-left corner of this figure). It also works nicely in finding multiple sources overlapping within their source radii (e.g., source no. 30 and 33, 32 and 42, in Figure 2 – see section 3.3 on their photometry). One known exception (not specific to **wavdetect**) is a problem due to the PSF shape, which is not circularly symmetric. On a rare occasion, a pair of spurious sources may be detected at the location of a single source when the source is far enough off-axis and bright enough to manifest the PSF shape. Figure 7 illustrates this effect – an observed image (the left panel) of a double-peaked source with ~2000 counts at an off-axis distance of 6′ is compared with a single PSF image generated (the right panel) at the source location. In this case, 2 sources are detected by **wavdetect**, 1.6″ apart. Also plotted in Figure 7 is a Richardson-Lucy de-convolved image (the bottom-left panel) confirming that this is really a single source. To identify this PSF effect, we have inspected source pairs with small separation. We have found 3 pairs of sources affected. The correct positions were re-determined by the PSF de-convolution and the extra sources were flagged (flag=015 in Table 3) in the database to exclude false sources in any further analysis.

**wavdetect** also provides source information such as a source count rate and size, but they may not be reliable, particularly when there is a nearby (extended) source. We rely on **wavdetect** results only for positions of detected sources (see section 5 for more discussions about the positional error) and determine source properties independently as described in section 3.3.

3.3. Determination of Source Properties

Source counts were extracted within a circle centered on the **wavdetect**-determined source position and background counts are extracted locally in an annulus surrounding the source. We choose the source extraction radius to be a 95% encircled energy radius (at 1.5 keV) as a function of off-axis angle (determined from the psfsize table, psfsize_2000830.fits, available in the CIAO and CALDB public distribution) with a minimum of 3 arcsec near the aim point and a maximum of 40 arcsec at the far edge of the field of view. Similarly, the background was estimated for each source in an annulus from 2 to 5 times source radius. When nearby sources exist within the background region, they are excluded before measuring the background count. Net count rates are then calculated with the effective exposure (including vignetting) for both the source and background regions. Errors are derived following Gehrels (1986).

When the source extraction regions of nearby sources overlap, their source count rates will be overestimated. Note that they are flagged in the database (see Table 3). To correct this, we have applied two independent methods. The first method calculates the source counts from a pie-sector, which excludes a nearby source region, and then rescales it based on the area ratio of the chosen pie to the full disk. Once the correction factor is determined, the same factor can be applied to correct counts in all energy bands. The $2^{nd}$ method is to fit a 2-D image with two (or more) PSFs generated at the source position. The advantage of the $2^{nd}$ method is to use all the photons to maintain the highest statistics. Its disadvantages, however, are (1) the 2-D fitting is less reliable when sources are faint, and (2) the whole task (generating a PSF and fitting) must be done in each energy band. The source counts corrected by the $1^{st}$ method are used in this paper.

Table 2. Energy bands and Definition of X-ray Colors

```
Energy band selection:
    Broad (B):      0.3 - 8.0 keV
    Hard (H):       2.5 - 8.0 keV
    Soft (S):       0.3 - 2.5 keV
    Soft1 (S₁):     0.3 - 0.9 keV
    Soft2 (S₂):     0.9 - 2.5 keV

Hardness Ratio and X-ray Colors
    HR  = (H-S) / (H+S)
    C21 = -log(S₂) + log(S₁) = log (S₁/S₂)
    C32 = -log(H)  + log(S₂) = log (S₂/H)

Additional energy bands:
    Conventional Soft (S_C)  0.5-2.0 keV
    Conventional Hard (H_C)  2.0-8.0 keV
```

The energy range used in this study is between 0.3 and 8 keV. The upper limit was selected to reduce background events while still including hard X-rays which could lead to interesting sources (such as heavily absorbed AGN), because background particles dominate and the HRMA (High Resolution Mirror Assembly) effective area steeply decreases at energies higher than 8 keV (van Speybroeck et al. 1997). We have divided the counts into 3 energy bands (see Table 2) in order to construct 2 X-ray colors. The energy boundaries were selected (1) to optimize photon statistics by distributing a comparable number of photons to each band, but the largest number of photons to the middle band (S2) which is used in both X-ray colors, (2) to directly compare with previous ROSAT results (< 2.5 keV), and (3) to confine most soft X-ray lines seen in soft X-ray sources (e.g., stellar sources) to a single band, $S_1$ (< 0.9 keV). With the 3 energy bands, we define two X-ray colors, following the same convention as in optical colors such that higher numbers are redder/softer. They are also consistent with X-ray colors defined in earlier studies with the Einstein X-ray data (e.g., Kim et al., 1992). We note that the sense is opposite to a hardness ratio in which higher numbers are harder (e.g., Green et al. 2003). The main advantage of this selection of X-ray colors is to isolate 2 spectral parameters, intrinsic hardness and absorption so that one X-ray color mostly represents absorption and another color a spectral slope (if $N_H \lesssim 10^{22}$ cm$^{-2}$). We describe the average X-ray colors in our sample in Paper II. Finally, to the ChaMP catalog, we have added two more commonly used energy bands (0.5-2.0 keV) and (2.0-8.0 keV) to provide users with flexibility and allow them to directly compare with other results.

Source fluxes are determined by calculating the energy conversion factor (ECF – actually count rate to flux conversion factor) for each observation (and each chip), because the quantum efficiency (QE) of ACIS CCDs varies with time and the galactic value of $N_H$ varies from one pointing to another. To calculate ECFs, we assume a power-law emission model of $\Gamma_{ph} = 1.7$ and $\Gamma_{ph} = 1.4$ with absorption by galactic $N_H$ determined for each observation (Stark et al. 1992). Those parameters were selected to be consistent with other results (e.g., Hasinger et al. 1993; Brandt et al. 2001) for direct comparison (e.g., $\Gamma_{ph} = 1.7$ for the soft band and $\Gamma_{ph} = 1.4$ for the hard band, as used in Paper II). The QE degradation (see CXC Memo on Jul. 29, 2002; http://cxc.harvard.edu/cal/Acis/ Cal_prods/qeDeg/ index.html) is most significant at energies below 1 keV (or in the S-band). To correct the time-dependent QE degradation per observation, we have used **sherpa** available in CIAO v2.3 (http://cxc.harvard.edu/ciao) in conjunction with **corrarf** available in http:// cxc.harvard.edu /cal/ Acis/ Cal_prods/qeDeg/corrarf.tar.gz). We note that the S-band ECF varies by about 20% (for about 20 months spanning our sample) due to the QE degradation, while the H-band ECF remains almost constant. However, for extremely soft sources (such as super soft sources, e.g., in Di Stefano and Kong 2003), the correction could be even higher.

3.3.1. Source Extent

To identify an extended source and measure the source extent, we have generated radial profiles of individual sources and fitted them with a Gaussian profile and a β-model. Because the X-ray background rapidly increases with increasing energy, we have used S band images for this purpose. We then compared the measured Gaussian σ and core radius with the PSF size at a given off-axis distance. We initially identified extended sources with σ (″) > 1.5 x 90% EE radius (the limit was empirically selected after several iterations of trial and error). Then an individual source is re-checked for its extendedness (see section 3.5). This method works for most extended sources. However, it is difficult to determine extent for a faint source falling at a large off-axis distance and to model PSFs appropriate for sources at large off-axis angles by the simple Gaussian function we have applied. We have, therefore, limited our method to 4 CCDs (CCDID = 0-3) in ACIS-I observations and $D_{off-axis}$ < 10 arcmin in ACIS-S observations. Once a source is identified as extended, we then re-calculate its counts and flux, based on its size. Among the sources (3177 sources; see section 6) found in CCDID=0-3 in ACIS-I and CCDID=6-7 in ACIS-S observations, we have identified 21 extended (non-target) sources with Gaussian σ ranging from a few arcsec to 12 arcsec. In the first ChaMP catalog (991 sources; see section 6), we have 4 extended sources (flag=051 in Table 7).

We note that the detection probability of extended sources is more complex than that of point sources (in Section 4) and strongly depends on the background level as well as the source properties such as their flux and size. We will present a full simulation analysis for the extended source selection and completeness in the subsequent paper.

3.3.2. Variability Analysis

An investigation of the variability of sources detected in a survey such as ChaMP is complicated by the large ranges in observed source count rates and in the exposure times of the different ChaMP fields. Traditional methods based on light curve analyses pose difficulties for two main reasons. First, for a given source in the absence of significant background, the signal-to-noise ratio (SNR) in each bin varies as the square root of the bin length. The threshold for detection of variability is, then, always dependent on the bin size adopted for the light curve. Second, the time resolution of variability is limited to the Nyquist sampling of the light curve, and is equivalent to an interval of twice the adopted bin size. A general astrophysical X-ray source population that would be expected to contain AGN, early-type and late-type stars, X-ray binaries and X-ray pulsars, would exhibit timescales of variability ranging from a fraction of a second to days. There is clearly no obvious single binning scheme that could be optimized to encompass such a dynamic range in timescale. Alternative hierarchical binning methods that apply light curves multiple times in order to sample the full range of useful bin sizes are computationally expensive, and are not very sensitive to the type of variability characterized by generally quiescent behavior upon which might sit small, infrequent bursts.

To cope with the general X-ray source case, we have developed a variability test based on the ``Bayesian Blocks'' (BB) method of Scargle (1998). One advantageous property of X-ray detectors such as microchannel plates, proportional counters, and fast frame CCD cameras is that individual photon events are measured and timed (in the case of the latter, this is only true if the frame time is significantly smaller than the average time between events). Binning of time-tagged event data is unnecessary for examining time variability because event arrivals are described by the Poisson distribution. Deviations from the expected arrival times can then be exploited to investigate variability. The BB method is based on Bayesian statistics and seeks to determine the most probable segmentation of the observation into time intervals---``Bayesian Blocks''---during which the photon arrival rate has no statistically significant variations. The analysis method itself does not impose a lower limit to the timescale on which variability can be detected: this is instead determined by the timing accuracy of the observing instrument. The BB method has the further advantage that the Bayesian Blocks themselves describe the variability of any source in the most economical way using the minimum possible number of parameters, each parameter being the start and stop times of the block and the event rate during the block. The method is thus well-suited to having results for a large number of sources stored on a computer. Further, for each source divided into $n$ blocks, each block with a mean count rate $C_i$ and Poisson uncertainty $\sigma_i$, we can also usefully characterize the variability with three parameters: the median block length, the total source ``amplitude'', $\Phi$,

$$\Phi = \sum_{i=1}^{n-1} |C_{i+1} - C_i|$$

and the ``significance'' of variability, $S$,

$$S = \frac{\sum_{i=1}^{n-1} |C_{i+1} - C_i|}{\sqrt{\sigma_1^2 + 2\sigma_2^2 + ... + 2\sigma_{n-1}^2 + \sigma_n^2}}$$

One complication to the BB method as implemented in ChaMP is that the ACIS CCD camera upon which our survey is based has a typical frame time of 3.2s. Individual event time tags are therefore only accurate to ±1.6s. For faint sources where event arrivals are separated on average by intervals much larger than this, the frame time is irrelevant for source variability, except in the rare case of a repeating signal with a comparable or shorter period. However, for sources in which multiple events might arrive in the same frame, this poses a problem in that all events within the frame nominally arrive at exactly the same time. This artifact can give rise to spurious variability at the readout frequency as the BB algorithm perceives all the events as arriving at the end of a 3.2s interval. In order to avoid this, for frames in which multiple events occur we artificially distribute the events evenly in time throughout the frame. An example of the output of the BB analysis is shown in Figure 8. Among the sources (3177 sources; see section 6) found in CCDID=0-3 in ACIS-I and CCDID=6-7 in ACIS-S observations, we have identified 92 variable sources. In the first ChaMP catalog (991 sources; see section 6), we have 53 variable sources (flag=055 in Table 7).

One disadvantage of the BB method is that it is not easy to account for background subtraction. While generally low and often negligible, the background can vary quite strongly during an observation. Source variability is, however, easily separated from background variability by examining the BB analysis of an annular region surrounding each source region.

In order to diagnose the possible presence of significant periodic variability, we also compute power spectra for each source. While these power spectra are currently examined only by eye, in future implementations of our variability analysis we expect to undertake more thorough and automated searches for periodicity.

3.4. Absolute Position

The absolute positional accuracy of the Chandra observatory is about 1 arcsec, when processed with the most recent calibration data, as specified in http://cxc.harvard.edu/cal/ASPECT/celmon. In order to detect any unforeseen error, particularly in those data processed with old calibration data, and to provide improved celestial positional accuracy for ChaMP X-ray sources, we use optical observations and databases which are being compiled for the ChaMP (Green et al. 2003). For each ChaMP field, we obtain optical CCD images in the SDSS g′, r′ and i′ bands. Optical sources are extracted in these images using SExtractor (Bertin and Arnouts 1996) for each field, we astrometrically calibrate our optical coordinate system against the GSC 2.2 catalog (Bucciarelli et al. 2001), which is in turn astrometrically calibrated via the Astrographic Catalog/Tycho and the Tycho Catalog II to the ICRS reference frame. For each ChaMP field, the positions of X-ray sources are correlated to optical sources in the field, to correct the X-ray coordinate system from standard Chandra aspect processing. The field correction is then applied to each X-ray source position. This typically corrects the X-ray positions by less than ~1 arcsec. As the positional accuracy of sources detected by **wavdetect** depends on how well the PSF is sampled, the error increases with decreasing source counts and increasing off-axis distances (see section 5). Therefore, we use only sources with counts $> 20$ and $D_{\text{off-axis}} < 300''$ to fine-tune the astrometric solution for the Chandra image.

3.5. Verification and Validation (V&V)

All the data products have been reviewed and confirmed by more than one ChaMP scientist (1) to make sure that the data processing was done correctly and (2) to flag those sources with various special issues listed in Table 3. If there was a problem in processing, the data were re-processed and re-examined. The source flags with * in Table 3 are initially flagged by automatic XPIPE processing, then confirmed by V&V, whereas those without * are determined only by visual examination. The flags in Table 3 are divided into 4 sub-groups: false sources, questionable sources, valid sources but with uncertain properties and sources with special characteristics. Most of them are self-explanatory or references are given in the table. Flag=021 (spurious source) is rather subjective, but we have only one source in our initial sample. Flag=037 (pile-up) is usually for X-ray bright target sources. For the remaining sources (i.e., without this flag), the pile-up is always less than a few % near on-axis and lower at large off-axis distances. Flag=054 is for already known X-ray jets (either by previous X-ray missions or by radio data). Searching for new faint X-ray jets or close multiple sources (lens candidates) are some of ChaMP science goals.

Table 3. Source flags

```
False X-ray sources.
  011 false source by a hot pixel or by a bad bias value (Figure 3b)
  012 false source by a bad column (Figure 3a)
  013 false source along the readout direction of a very strong source
  014 false source by the FEP 0/3 problem
      (http://cxc.harvard.edu/ciao/caveats)
  015 double sources detected by the PSF effect (Figure 7)

X-ray sources – questionable
  021 Visual inspections found it as a spurious source.

Valid sources, but source properties may be subject to a large uncertainty
  *031 bad pixel/column exists within source extraction radius
  *032 nearby source exists within the source extraction circle
  *033 nearby source exists within the bkgd extraction annulus
  *034 source is found near an extended source
  *035 bkgd region overlaps with a nearby extended source
  *036 source near the edge of the chip
```

```
  037 pileup (see Chandra POG)
  038 uncertain source position by flag=015 (Figure 7)

Other cases
*051 source is extended
  052 same source in multiple observations
*053 target of observation
  054 X-ray jet
*055 variable source

(* flagged by the automatic pipeline)
```

## 4. Detection Probability

From a statistical perspective, **wavdetect** is a hypothesis test. At any given location, the null hypothesis that there is no source is checked against the observed signal, and the null is rejected if the signal cannot be obtained at some threshold probability as a fluctuation due to the background. The usefulness of a statistical test depends both on its ability to correctly accept the null hypothesis (i.e., minimize the false positives, the number of spurious sources; this is the so-called Type-I error and is quantified by the threshold significance of detection) as well as correctly reject the null (i.e., minimize the false negatives, the number of real sources that are missed; this is the so-called Type-II error and is quantified by the probability of detecting a source).

To quantitatively determine the performance of **wavdetect** (Kashyap et al. 2003, in preparation), we have run a series of simulations using MARX (MARX Technical Manual) and have also made use of Chandra data to confirm some of the results of the simulation. First, postage stamp (256 x 256) images are simulated at different off-axis locations (0', 2', 5', and 10') for a grid of source strengths (ranging from ~3 to ~4000 counts) and background intensities (ranging from ~1.5 x $10^{-4}$ count pixel$^{-1}$ to ~0.2 count pixel$^{-1}$), assuming a flat spectrum between 0.2 and 10 keV (the input spectrum is not important, because the only interesting parameter is the counts produced in the detector by the source and background). Such simulations were carried out ~50 times at each grid point. **wavdetect** was then run on each image adopting a detection threshold of 1 expected false source per image, at scales 1,2,4 pix for the on-axis points and 2,4,8,16 for the off-axis locations.

### 4.1. Type I Errors

We find that the performance of **wavdetect** is as expected (~1 false source is found in each image) for on-axis sources, and *improves* for off-axis sources. At 10' off-axis, the number of false sources is on average < 0.2 per simulation. The reason for the improvement is the additional logic discrimators installed in wrecon, the second part of **wavdetect**, which compare source sizes with the known PSF size and eliminates many false sources in that manner.

We have also performed a comparable study with a Chandra observation of relatively long exposure times (~100 ksec). We have split the long observation into smaller pieces (10 ksec each) and run **wavdetect** on the original long observation and smaller segments with the same parameters. It would be reasonable to assume that sources found in the segments as well as in the original observation are likely to be real whereas sources found in the segments but not in the original observation are likely to be false. We can then measure how many spurious sources are found in the segments, but not found in the original, long observation. We have performed this exercise with obsid=536 and obsid=927. On average, we found 0.3 spurious sources per CCD, which is fully consistent with the simulation results.

4.2. Type II Errors

Determining the rate of false negatives, or the probability of failing to detect a real source, is more complex because of its strong dependence on source strength, background intensity, and off-axis location. The detection probability decreases as the background rate increases, because it is easier to obtain fluctuations from the background that match the source intensities, thus reducing the significance of detection. Also, in general the PSF increases in size at larger off-axis angles, spreading the source counts over a larger area of the detector, thus including a larger number of background counts within the source region, which again serves to reduce the detection probability. Finally, as the nominal source strength is reduced, the Poisson fluctuations in the observed source counts ensure that in increasingly larger numbers, sources lie below the detection thresholds, again reducing the probability of detecting the source.

We determine the detection probability over the same range of parameters as above, and show representative curves in Figure 9. It is clear that Type-II errors are critical at the faint end, i.e., near the detection limit of each observation. For instance, only half the sources with a strength of say 10 counts are detected at 5' off-axis when the background is ~0.03 count pixel$^{-1}$.

5. Positional Uncertainty

With the superb Chandra spatial resolution, the on-axis positional accuracy is expected to be accurate with an error less than 1″. However, with increasing off-axis angle, PSFs spread out and also become circularly asymmetric (see Chandra POG). Consequently, a source position may not be reliable for a weak source at a large off-axis angle. In order to quantify this uncertainty, we have carried out another set of simulations with SAOSAC (http://cxc.harvard. edu/chart) because its ray trace technique represents the actual HRMA more realistically than MARX. Figure 10 illustrates an example of simulated sources in 4 ACIS-I CCDs. We ran **wavdetect** on the simulated images and compared **wavdetect**-determined source positions with input positions. Figure 11 shows the positional error as a function of off-axis angle for a wide range of source strengths, from 20 to 10,000 counts. When a source lies at the edge of the CCD, source photons may be lost and the **wavdetect** centroiding algorithm may not work properly. Such sources (with average exposure within the source radius < 80% of the on-axis exposure) are marked as a plus symbol in Figure 11. In the following discussion, we exclude those sources falling near the CCD edge. We note that these sources are flagged in the ChaMP database to indicate their large uncertainties both in flux and position. As expected, the source position is accurately determined for strong sources with 1,000 counts and 10,000 counts (Figure 11-c and d). The positional uncertainty is ~0.7″ (95% confidence), regardless of $D_{off-axis}$. For fainter sources (with 20 counts and 100 counts in Figure 11-a and b, respectively), the positional error remains relatively small (less than 1-2 arcsec) within $D_{off-axis} < 6'$. However, the positional error increases significantly at $D_{off-axis} > 6'$. The results are summarized in Table 4, where a mean value, 1σ scatter from the mean, positional uncertainties at 67, 95 and 99% confidence levels are listed.

By examining sources with large offsets between the simulated and detected position, we have recognized that source positions estimated by **wavdetect** (in CIAO 2.3 or earlier) or the algorithm given in Freeman et al. (2002) become less accurate by up to several arcsec (see Figure 11), if either the contribution of the background to the observed total counts in the source cell is non-negligible, or the source cell is highly asymmetric. To mitigate this problem, the algorithm has been enhanced (Freeman, private communication). First, a position estimate is made using the original algorithm, along with an estimate of the error. Then, if the source cell size is at least 15 pixels, and if the first position estimate is at least three σ away from the nearest maximum in the source counts image, the position estimate is refined by the following method. Any asymmetry in the source cell shape, or asymmetry in the distribution of data around the true centroid, can lead to inaccuracies in the position determination, since outlier data will "pull" the estimated centroid away from its true position. (This effect is thus most noticeable far off-axis, for low counts, as demonstrated in Figure 11.) Taking the (possibly asymmetric) source cell and the original position ($x_o$, $y_o$) as an input, we eliminate the systematic effect of an asymmetric source cell by creating a new, temporary source cell: the largest square box centered at ($x_o$, $y_o$) that fits within the original source cell (i.e., all pixels in the new cell must also be within the old cell, but not vice-versa). A new centroid determination is made within this cell. Refinement is done iteratively, i.e., the process of creating a cell and centroiding is repeated until the refined position estimate stabilizes. Figure 12 illustrates this problem and its correction. The larger (red) ellipses indicate the source regions determined by a single run of **wavdetect**, while the smaller (blue) ellipses indicate the refined source regions, whose centroids are clearly closer to the actual photon distribution. In this example (obsid=520), the two sources with ~60 counts each are located at $D_{off-axis}$ ~ 8 – 9′ and the corrections are 4.4″ (lower one) and 2.5″ (upper one).

Table 4

Positional error (in ″) at $D_{off-axis}$ = 6-8′ for count = 20

| Method | number of sources | mean error | σ | 67% | 95% | 99% |
|---|---|---|---|---|---|---|
| wavdetect in ciao 2.3 | 403 | 1.28 | 0.74 | 1.53 | 2.75 | 3.36 |
| Revised wavdetect | 403 | 1.13 | 0.64 | 1.34 | 2.31 | 3.17 |

Positional error (in ″) at $D_{off-axis}$ = 8-10′ for count = 20

| Method | number of sources | mean error | σ | 67% | 95% | 99% |
|---|---|---|---|---|---|---|
| wavdetect in ciao 2.3 | 146 | 2.82 | 1.8 | 3.46 | 5.82 | 7.66 |
| Revised wavdetect | 146 | 1.85 | 1.2 | 2.11 | 4.23 | 4.92 |

The positional error after this refinement decreases considerably, in particular for a faint source at a large off-axis angle (see Figure 13 and Table 4). In summary, the positional error is 2″ or better for a source (regardless of its strength) within $D_{off-axis}$ < 6′ from the aim point, while the error for a faint source increases to 2-3″ and 4-5″ at $D_{off-axis}$ = 6-8′ and 8-10′, respectively (95% confidence). We have reexamined by eye those sources with large errors and confirmed that the error is basically driven by the statistical noise with a small number of source photons spreading out inside a large PSF. In XPIPE processing, we apply the improved position determination algorithm to all sources at $D_{off-axis}$ > 400″ and used the improved position and error in the following discussions (section 6).

A set of empirical formulae (applicable up to $D_{off-axis}$ < 10′) are given here for a quick, approximate estimate of an error box:

$$PE (″) = 1 - 0.02\ D_{off-axis}^2 + 0.0067\ D_{off-axis}^3 \text{ for sources with 20 counts}$$
$$\phantom{PE (″) =} 1 - 0.01\ D_{off-axis}^2 + 0.0025\ D_{off-axis}^3 \text{ for sources with 100 counts}$$

where $D_{off-axis}$ is in unit of arcmin. In the ChaMP source catalog, we provide the positional error for an individual source by interpolating these formulae and applying a conservative minimum of 1″ (see Section 3.4 and cxc.harvard.edu/cal/ASPECT/celmon for Chandra absolute astrometric accuracy).

Given that the PSF is asymmetric and the direction of PSF elongation is a function of azimuth, we have searched for any systematic trend, such as a preferential direction of offsets (i.e., both radial and tangential offsets) as a function of azimuth. However, we do not see any significant trend, as both the amount and direction of offsets appear to be random on a scale of < 1 arcsec. It is possible there still remains a systematic effect on a smaller scale (~ a few tenth arcsec).

We have also performed a comparable study with Chandra data. Utilizing multiple observations pointing to the same part of the sky, we compare the multiple source positions of the same source measured in different observations. For this purpose, we have selected the Chandra Deep Field North (CDF-N) data which consist of 12 separate observations from Nov. 1999 to Mar. 2001. Excluding observations with a relatively short exposure time, we have used 10 observations with exposure times ranging from 50 ksec to 170 ksec. We ran **wavdetect** on each observation and cross-correlated source positions. To remove the systematic shift between different observations by an absolute position error, we first cross-correlate sources within 300″ from the aim point for which the positional accuracy is expected to be good as indicated by the above simulation study. We found a systematic shift of up to 2″ from one field to another. This is consistent with the known error in aspect calibration files of ACIS-I (http://cxc.harvard.edu/cal/ASPECT/). After registering each field to a common frame, we cross-correlated again all sources in 10 observations. The position difference of the same source detected in different observations is plotted against off-axis angle in Figure 14. In each matching pair, the larger off-axis angle was used here. The result is in full accordance with the simulation result. While most sources can be found within 1-2″, the positional error can be as large as 5″ at $D_{off-axis} > 6′$.

We note that the refined position does not affect source counts. The change using the revised position remains well within the error (Section 3), because the source extraction radius is much larger than the position change, when the correction is necessary, i.e., at the large $D_{off-axis}$.

6. ChaMP X-ray source catalog

62 fields (listed in Table 1) have been completed in XPIPE processing and follow-up manual V&V. We have found 4517 sources, after excluding false sources (flag=011- 021 in Table 3). Further excluding the target of each observation (flag=053), sources at the edge of CCDs (flag=036) and sources affected by pile-up (flag=037; in our 62 fields they all happened to be targets), we ended up with 4005 sources. 3177 sources are within CCDID=0-3 in ACIS-I observations and CCDID=6-7 in ACIS-S observations. Among these sources, we have used 991 X-ray sources to extract the first ChaMP results on X-ray source properties in terms of Log(N)-Log(S) and X-ray colors (see paper II). They are all near on-axis and bright X-ray sources, which were selected (1) to avoid various systematic effects (e.g., detection probability, positional error) and (2) to maintain relatively high statistical significance (see Table 5). We present here the first ChaMP source catalog consisting of these 991 sources. Figure 15 (a-d) show the distributions of exposure times and net source counts in 3 energy bands for all sources in 62 fields and the 991 sources listed in the catalog.

Table 5 Selection Criteria for the first ChaMP catalog

|  | no. of sources | Counts limit | Off-axis distance limit |
|---|---|---|---|
| S-band Log(N)-Log(S) | 707 | 20 in S-band | 400" in ACIS-I or S3 in ACIS-S |
| H-band Log(N)-Log(S) | 236 | 20 in H-band | 400" in ACIS-I or S3 in ACIS-S |
| X-ray colors | 620 | 50 in B-band | I0-I3 in ACIS-I or S2-S3 in ACIS-S |

To facilitate the use of the ChaMP catalog, we provide ECFs for various energy bands and various spectral models in Table 6. As described in section 3.3, ECFs are calculated for different observation times (and per chip) and galactic $N_H$ for each field. In combination with Table 8 (containing photometric information), users can easily generate necessary fluxes. Due to the limited space, we present the only abbreviated version of the Catalog in Table 7 (source position and flag), 8 (source photometric information) and 9 (source spectral information). The full tables are available in the electronic version of this paper and also in the ChaMP web site, http://hea-www.cfa.harvard.edu /CHAMP. Descriptions of each column in Table 7-9 are listed below.

Table 7:
Col (1) ChaMP source id given by ra and dec in J2000.
Col (2) obsid
Col (3) ccdid (0-3 for ACIS-I and 4-9 for ACIS-S)
Col (4) source id number assigned by **wavdetect**
Col (5-6) RA and DEC (J2000) after refinement in degree (note. Col (1) in hms) (see section 5)
Col (7) positional error in arcsec (see section 5).
Col (8) off-axis distance in arcmin
Col (9) source extraction radius (95% EE radius; see section 3.3)
Col (10) effective exposure in ksec (see section 3.1)
Col (11) source flag (see section 3.5)

Table 8:
Col (1) same as Col (1) in Table 7
Col (2-3) net counts in the B-band (0.3- 8.0 keV) and errors (see section 3.3)
Col (4-5) net counts in the $S_1$-band (0.3- 0.9 keV) and errors (see section 3.3)
Col (6-7) net counts in the $S_2$-band (0.9- 2.5 keV) and errors (see section 3.3)
Col (8-9) net counts in the H-band (0.3- 8.0 keV) and errors (see section 3.3)
Col (10-11) net counts in the $S_C$-band (0.5- 2.0 keV) and errors (see section 3.3)
Col (12-13) net counts in the $H_C$-band (2.0- 8.0 keV) and errors (see section 3.3)
Col (14) flux in the B-band (0.3-8.0 keV) (see section 3.3 and Table 6)

Table 9:
Col (1) same as Col (1) in Table 7
Col (2-3) hardness ratios and errors (see section 3.3)
Col (4-5) C21 and errors (see section 3.3)
Col (6-7) C32 and errors (see section 3.3)
Note that in both colors, we assign 99 for an upper limit, –99 for a lower limit, and 999 for undetermined.

7. Conclusion

(1) We present the first ChaMP X-ray source catalog of 991 sources after applying uniform data reduction techniques. The source properties include photometry (in various energy bands), spectroscopy (using hardness ratio and X-ray colors), spatial extendedness and time variability.

Careful tests and simulations of the ChaMP XPIPE X-ray data processing described in this paper yield the following results:

(2) The type I error (detecting a false source) is always less than 1 per CCD with the selected parameter, threshold=$10^{-6}$.

(3) The type II error (missing a real source) is a complicated function of source strength, background level and off-axis distance. The error could be as large as 50% for a weak source (with ~10 photons) at a moderate off-axis distance ($D_{off-axis}$ ~ 5′) with a typical background rate (~0.03 count pixel$^{-1}$) and should be carefully incorporated in any statistical analysis.

(4) The positional accuracy is always good (< 1″) for a bright source, regardless of its off-axis distance. However, for a faint source at a large off-axis distance, the position uncertainty can be as high as 4-5 arcsec (95% confidence).

(5) We have developed new tools to identify extended or variable sources. In particular, our new variability test, based on the Bayesian Blocks algorithm, is applicable to common, faint sources.

This work has been supported by CXC archival research grant AR2-3009X. We acknowledge support through NASA Contract NAS8-39073 (CXC). We thank the CXC DS and SDS teams for their supports in pipeline processing and data analysis.

Figure Captions

Figure 1: ChaMP predicted (a) number of sources and (b) effective sky area. Predictions are determined by simulations for 137 ChaMP fields, based on a deep Log(N)-Log(S). Also included are analogous simulations for the combined CDFs (2 Msec North and 1 Msec South) and for the ROSAT surveys analyzed by Miyaji et al (2000).

Figure 2. An example XPIPE output image. A circle indicates an X-ray source detected by **wavdetect** and its radius is proportional to the PSF size at the off-axis distance for a given source. 45 sources are detected, including one extended source (a target source in this observation) and a few overlapping sources.

Figure 3. Examples of (a) a bad column and (b) a hot pixel in sky coordinates. (a) A series of false sources are detected along the bad column. (b) Two hot pixels are seen in a Lissajous pattern (near the center) and 4 false sources each are detected. In ChaMP data processing, the presence of unfiltered bad columns and bad pixels is checked by visual examination of images in chip coordinates.

Figure 4. A sample image on S4 chip (ccdid=8) before (bottom) and after (top) the de-streaking correction.

Figure 5. Comparison of background light curves of (a) BI S3 (ccdid=7) chip and (b) FI S2 (ccdid=6) chip. Note that they were made from the same observation. Time intervals during which the background rate is high (beyond 3 sigma) are marked by a red cross.

Figure 6. Histogram of the effective exposure times for BI and FI chips. Background flares significantly reduce the effective exposure time in the BI chip.

Figure 7. Double peaked sources due to the PSF. The top-left panel is the observed image of a single source with ~2000 counts at $D_{off\text{-}axis}=6'$ and the top-right panel is the PSF image generated at the source location. In this case, 2 sources are detected by **wavdetect**, 1.6″ apart. The bottom-left panel clearly shows a single source after applying the Richardson-Lucy deconvolution.

Figure 8. (a) An example of variability determined by the Bayesian Blocks method. The Chandra data of 1WGA J1216.9+3347 is used here to illustrate the results of our variability analysis. The dotted histogram represents the light curve and the thick black histogram indicates where the count rate remains constant or varies. The break in the count rate is also marked by the red vertical line. (b) The light curve of the same source determined by the traditional time binning method (taken from Cagnoni et al. 2003).

Figure 9. Detection probability as a function of background counts with various source counts (from a few to ~100, indicated at the right side of figures) and off-axis distances: (a) on-axis, (b) 2′ off-axis, (c) 5′ off-axis and (d) 10′ off-axis.

Figure 10. SAOSAC simulation of point sources at a wide range of off-axis angles in 4 ACIS-I chips. Each source has 1000 net input counts.

Figure 11. **wavdetect** position errors measured with SAOSAC simulations. About 2000 sources are simulated each for 20, 100, 1000 counts (a-c), and about 200 sources for 10000 counts (d). Differences between estimated and expected positions are plotted against off-axis angle. Sources falling at the detector edges are subject to a large error and are denoted by +. For visibility, points are horizontally shifted by adding random numbers (up to 50 arcsec) to off-axis angles.

Figure 12. Examples of X-ray sources identified by **wavdetect** with relatively large position errors. The large ellipses are the source regions determined by **wavdetect** in CIAO 2.3. Note that the centers of two ellipses (marked by x) are off by 2-4″ from the local peaks. The smaller ellipses indicate the revised source regions by the new algorithm described in section 5.

Figure 13. Same as Figure 11 after the position correction with a new wavdetect algorithm.

Figure 14. Position error measured in 10 observations of Chandra Deep Field-North. The position difference of the same source is plotted against off-axis angle for which we take the larger one in each pair. Sources which lie at the detector edges at least in one observation of the matching pair are denoted by +.

Figure 15. Distribution of exposure times and source counts in 3 energy bands.
(a) Effective exposure time after correcting for the CCD dead time and background flares. (b-d) Net counts of 3177 sources obtained in 62 fields (solid histograms) and of 991 sources used in the 1$^{st}$ ChaMP catalog as described in Table 5 (dashed histogram).

Table 1. List of Chandra Fields used in this paper

| Obsid | Seq no | Target | RA (2000) | DEC (2000) | N(H) ($10^{20}$cm$^{-2}$) | obs. date | Exp (ksec) | aim point detector | CCD used (*) |
|---|---|---|---|---|---|---|---|---|---|
| 520 | 800028 | MS 0015.9+1609 | 0 18 32.7 | + 16 30 4.0 | 4.06 | Aug 18, 2000 | 60.99 | ACIS-I | 01236 |
| 2242 | 900069 | GSGP4X:048 | 0 57 17.9 | − 27 22 23.8 | 1.69 | Dec 18, 2000 | 6.66 | ACIS-S | 235678 |
| 2244 | 900071 | GSGP4X:069 | 0 57 38.9 | − 27 33 30.0 | 1.62 | Oct 30, 2000 | 6.86 | ACIS-S | 235678 |
| 2245 | 900072 | GSGP4X:082 | 0 57 51.9 | − 27 23 30.6 | 1.43 | Oct 30, 2000 | 6.52 | ACIS-S | 235678 |
| 2247 | 900074 | GSGP4X:109 | 0 58 26.4 | − 27 29 51.0 | 1.37 | Nov 07, 2000 | 10.88 | ACIS-S | 235678 |
| 2248 | 900075 | GSGP4X:114 | 0 58 38.3 | − 27 49 17.5 | 1.55 | Nov 08, 2000 | 10.11 | ACIS-S | 235678 |
| 521 | 800029 | CL 0107+31 | 1 2 5.2 | + 31 47 54.7 | 5.49 | Oct 23, 1999* | 46.87 | ACIS-I | 01237 |
| 342 | 700014 | NGC 526A | 1 23 53.6 | − 35 4 33.5 | 2.09 | Feb 07, 2000 | 5.78 | ACIS-S | 235678 |
| 913 | 800089 | CL J0152.7-1357 | 1 52 49.3 | − 13 56 19.1 | 1.61 | Sep 08, 2000 | 34.81 | ACIS-I | 012367 |
| 1642 | 700258 | HE0230-2130 | 2 32 35.7 | − 21 17 12.2 | 2.27 | Oct 14, 2000 | 8.35 | ACIS-S | 123678 |
| 525 | 800033 | MS 0302.7+1658 | 3 5 28.3 | + 17 13 20.6 | 10.95 | Oct 03, 2000 | 8.95 | ACIS-I | 01236 |
| 796 | 600099 | SBS 0335-052 | 3 37 44.5 | − 5 2 19.4 | 4.98 | Sep 07, 2000 | 46.81 | ACIS-I | 012367 |
| 624 | 200049 | LP 944-20 | 3 39 36.8 | − 35 26 21.2 | 1.44 | Dec 15, 1999* | 40.94 | ACIS-S | 23678 |
| 902 | 800078 | MS 0451.6-0305 | 4 54 12.9 | − 2 58 52.4 | 5.18 | Oct 08, 2000 | 41.53 | ACIS-S | 235678 |
| 346 | 700018 | PICTOR A | 5 19 45.6 | − 45 46 29.3 | 4.12 | Jan 18, 2000* | 25.44 | ACIS-S | 23678 |
| 914 | 800090 | CL J0542.8-4100 | 5 42 49.0 | − 40 58 48.4 | 3.59 | Jul 26, 2000 | 48.72 | ACIS-I | 01236 |
| 377 | 700049 | B2 0738+313 | 7 41 11.9 | + 31 12 35.8 | 4.18 | Oct 10, 2000 | 26.91 | ACIS-S | 235678 |
| 838 | 700143 | 3C 200 | 8 27 26.8 | + 29 19 19.9 | 3.69 | Oct 06, 2000 | 10.07 | ACIS-S | 235678 |
| 1643 | 700259 | APM08279+5255 | 8 31 43.9 | + 52 45 48.7 | 3.91 | Oct 11, 2000 | 6.87 | ACIS-S | 123678 |
| 2130 | 700320 | 3C 207 | 8 40 49.2 | + 13 12 57.0 | 4.14 | Nov 04, 2000 | 22.90 | ACIS-S | 235678 |
| 1708 | 800103 | CL 0848.6+4453 | 8 48 54.7 | + 44 54 33.3 | 2.73 | May 03, 2000 | 59.39 | ACIS-I | 012367 |
| 927 | 800103 | CL 0848.6+4453 | 8 48 54.8 | + 44 54 32.9 | 2.73 | May 04, 2000 | 122.18 | ACIS-I | 012367 |
| 1596 | 700212 | 0902+343 | 9 5 32.8 | + 34 9 7.9 | 2.28 | Oct 26, 2000 | 9.68 | ACIS-S | 235678 |
| 2227 | 800166 | RX J0910+5422 | 9 10 39.7 | + 54 19 54.8 | 1.98 | Apr 29, 2001 | 104.25 | ACIS-I | 01236 |
| 419 | 700091 | RX J0911.4+0551 | 9 11 26.8 | + 5 50 57.3 | 3.70 | Nov 02, 1999* | 24.55 | ACIS-S | 01237 |
| 1629 | 700245 | RXJ0911.4+0551 | 9 11 28.7 | + 5 51 25.9 | 3.70 | Oct 29, 2000 | 9.13 | ACIS-S | 123678 |
| 839 | 700144 | 3C 220.1 | 9 32 35.0 | + 79 7 10.8 | 1.90 | Dec 29, 1999* | 17.14 | ACIS-S | 23678 |
| 805 | 600108 | I ZW 18 | 9 33 56.5 | + 55 14 37.7 | 1.99 | Feb 08, 2000 | 24.49 | ACIS-S | 235678 |
| 926 | 800102 | MS 1008.1-1224 | 10 10 14.7 | − 12 41 4.9 | 6.74 | Jun 11, 2000 | 43.87 | ACIS-I | 012367 |
| 512 | 800020 | EMSS 1054.5-0321 | 10 56 55.8 | − 3 39 20.3 | 3.67 | Apr 21, 2000 | 75.60 | ACIS-S | 123678 |
| 915 | 800091 | CL J1113.1-2615 | 11 12 54.1 | − 26 15 41.2 | 5.52 | Aug 13, 2000 | 101.35 | ACIS-I | 012367 |
| 363 | 700035 | PG 1115+080 | 11 18 15.1 | + 7 45 16.0 | 4.01 | Jun 02, 2000 | 24.42 | ACIS-S | 123678 |
| 1630 | 700246 | PG1115+080 | 11 18 18.5 | + 7 46 29.8 | 4.01 | Nov 03, 2000 | 9.73 | ACIS-S | 123678 |
| 868 | 700173 | PG 1115+407 | 11 18 42.8 | + 40 25 17.6 | 1.91 | Oct 03, 2000 | 17.37 | ACIS-I | 012367 |
| 2126 | 700316 | 3C263 | 11 40 5.2 | + 65 47 59.7 | 1.15 | Oct 28, 2000 | 29.15 | ACIS-S | 235678 |
| 898 | 800074 | B1138-262 | 11 40 46.1 | − 26 30 20.9 | 4.96 | Jun 06, 2000 | 23.50 | ACIS-S | 235678 |
| 536 | 800044 | MS 1137.5+6625 | 11 40 47.1 | + 66 7 19.7 | 1.18 | Sep 30, 1999* | 114.61 | ACIS-I | 012367 |
| 1712 | 790054 | 3C 273 | 12 29 6.3 | + 2 3 14.0 | 1.79 | Jun 14, 2000 | 12.04 | ACIS-S | 456789 |
| 325 | 800063 | S-Z CLUSTER | 13 12 22.4 | + 42 41 42.8 | 1.37 | Dec 03, 1999* | 80.61 | ACIS-S | 23678 |
| 2228 | 800167 | RX J1317.4+2911 | 13 17 12.2 | + 29 10 18.3 | 1.04 | May 04, 2001 | 108.09 | ACIS-I | 01236 |
| 809 | 700114 | MRK 273X | 13 44 43.0 | + 55 54 16.4 | 1.09 | Apr 19, 2000 | 40.93 | ACIS-S | 012367 |
| 507 | 800015 | RX J1347-114 | 13 47 28.7 | − 11 46 24.2 | 4.88 | Apr 29, 2000 | 9.90 | ACIS-S | 235678 |
| 1588 | 700204 | 3C294 | 14 6 50.4 | + 34 11 20.0 | 1.21 | Oct 29, 2000 | 19.02 | ACIS-S | 235678 |
| 578 | 890023 | 3C295 | 14 11 11.5 | + 52 13 1.6 | 1.34 | Aug 30, 1999* | 15.80 | ACIS-S | 235678 |
| 930 | 800106 | H1413+117 | 14 15 43.9 | + 11 30 0.1 | 1.80 | Apr 19, 2000 | 24.09 | ACIS-S | 456789 |
| 541 | 800049 | V1416+4446 | 14 16 43.0 | + 44 48 28.5 | 1.24 | Dec 02, 1999* | 29.67 | ACIS-I | 01236 |
| 907 | 800083 | QB 1429-008A,B | 14 32 29.5 | − 1 5 58.4 | 3.54 | Mar 31, 2000 | 21.32 | ACIS-I | 01236 |
| 869 | 700174 | ARP 220 | 15 34 54.7 | + 23 29 52.5 | 4.29 | Jun 24, 2000 | 54.18 | ACIS-S | 235678 |
| 326 | 800064 | 3C 324 | 15 49 46.3 | + 21 25 19.3 | 4.31 | Jun 25, 2000 | 31.95 | ACIS-S | 235678 |
| 546 | 800054 | MS 1621.5+2640 | 16 23 25.4 | + 26 36 12.4 | 3.59 | Apr 24, 2000 | 29.57 | ACIS-I | 01236 |
| 615 | 200040 | VB 8 | 16 55 34.0 | − 8 24 7.6 | 13.39 | Jul 10, 2000 | 8.54 | ACIS-S | 456789 |
| 548 | 800056 | RX J1716.9+6708 | 17 17 1.0 | + 67 11 44.1 | 3.71 | Feb 27, 2000 | 50.35 | ACIS-I | 01236 |
| 841 | 700146 | 3C 371 | 18 6 52.3 | + 69 50 5.5 | 4.84 | Mar 21, 2000 | 9.43 | ACIS-S | 456789 |
| 830 | 700135 | JET OF 3C 390.3 | 18 41 46.8 | + 79 48 21.2 | 4.16 | Apr 17, 2000 | 22.71 | ACIS-S | 235678 |
| 551 | 800059 | MS 2053.7-0449 | 20 56 18.6 | − 4 34 32.3 | 4.96 | May 13, 2000 | 42.34 | ACIS-I | 01236 |
| 928 | 800104 | MS 2137.3-2353 | 21 40 14.8 | − 23 40 22.0 | 3.57 | Nov 18, 1999* | 29.09 | ACIS-S | 23678 |
| 1644 | 700260 | HE2149-2745 | 21 52 7.8 | − 27 32 28.2 | 2.33 | Nov 18, 2000 | 9.18 | ACIS-S | 123678 |
| 1479 | 980429 | LEONID ANTI-RADIANT | 22 13 12.7 | − 22 10 43.4 | 2.49 | Nov 17, 1999* | 20.02 | ACIS-I | 01236 |
| 789 | 600092 | HCG 92 | 22 35 58.5 | + 33 59 31.4 | 7.74 | Jul 09, 2000 | 19.60 | ACIS-S | 235678 |
| 431 | 700103 | EINSTEIN CROSS | 22 40 27.9 | + 3 21 19.2 | 5.34 | Sep 06, 2000 | 21.89 | ACIS-S | 123678 |
| 918 | 800094 | CL J2302.8+0844 | 23 2 47.4 | + 8 45 14.7 | 5.05 | Aug 05, 2000 | 106.09 | ACIS-I | 012367 |
| 861 | 700166 | Q2345+007 | 23 48 18.1 | + 0 58 36.4 | 3.81 | Jun 27, 2000 | 65.00 | ACIS-S | 123678 |

- CCDID=0-3 for ACIS-I and 4-9 for ACIS-S (Chandra Proposer's Observatory Guide)
- N$_H$ from Stark et al. (1992)
- * CCD temperature > − 120 degree

Table 6 Energy Conversion Factors

| obsid | ccdid | gamma=1.2 B | S | H | Sc | Hc | gamma=1.4 B | S | H | Sc | Hc | gamma=1.7 B | S | H | Sc | Hc |
|---|---|---|---|---|---|---|---|---|---|---|---|---|---|---|---|---|
| 325 | 6 | 116.45 | 38.88 | 297.53 | 44.69 | 240.97 | 104.69 | 39.35 | 290.50 | 45.04 | 229.90 | 91.56 | 39.99 | 281.82 | 45.70 | 214.83 |
| 325 | 7 | 86.98 | 26.92 | 277.71 | 32.84 | 219.52 | 75.15 | 26.38 | 270.16 | 32.50 | 208.41 | 61.56 | 25.40 | 260.76 | 32.04 | 193.39 |
| 326 | 6 | 123.36 | 43.12 | 283.04 | 49.21 | 233.16 | 112.55 | 44.04 | 276.97 | 49.91 | 223.31 | 100.72 | 45.44 | 269.63 | 51.16 | 209.91 |
| 326 | 7 | 98.72 | 31.82 | 277.54 | 37.77 | 221.05 | 87.14 | 31.73 | 269.70 | 37.73 | 209.76 | 73.95 | 31.46 | 259.86 | 37.75 | 194.49 |
| 342 | 6 | 117.20 | 40.09 | 282.38 | 45.86 | 232.29 | 106.07 | 40.69 | 276.30 | 46.32 | 222.43 | 93.68 | 41.55 | 268.92 | 47.17 | 209.00 |
| 342 | 7 | 90.94 | 28.55 | 276.82 | 34.57 | 220.12 | 79.19 | 28.16 | 268.97 | 34.34 | 208.83 | 65.72 | 27.41 | 259.11 | 34.06 | 193.54 |
| 346 | 6 | 122.51 | 41.72 | 298.22 | 47.81 | 241.89 | 110.98 | 42.47 | 291.19 | 48.37 | 230.84 | 98.30 | 43.59 | 282.58 | 49.38 | 215.79 |
| 346 | 7 | 94.58 | 30.01 | 278.56 | 35.77 | 220.52 | 82.87 | 29.75 | 271.01 | 35.60 | 209.41 | 69.45 | 29.19 | 261.63 | 35.39 | 194.39 |
| 363 | 6 | 122.54 | 42.70 | 283.05 | 48.74 | 233.10 | 111.67 | 43.58 | 276.97 | 49.41 | 223.25 | 99.76 | 44.91 | 269.63 | 50.60 | 209.83 |
| 363 | 7 | 97.70 | 31.38 | 277.48 | 37.33 | 220.96 | 86.08 | 31.25 | 269.64 | 37.27 | 209.66 | 72.84 | 30.90 | 259.80 | 37.24 | 194.39 |
| 377 | 6 | 124.55 | 43.72 | 283.21 | 49.88 | 233.33 | 113.80 | 44.70 | 277.15 | 50.63 | 223.48 | 102.11 | 46.23 | 269.80 | 51.97 | 210.07 |
| 377 | 7 | 100.25 | 32.48 | 277.54 | 38.43 | 221.12 | 88.72 | 32.47 | 269.71 | 38.45 | 209.84 | 75.61 | 32.30 | 259.88 | 38.54 | 194.58 |
| 419 | 7 | 91.55 | 28.76 | 278.33 | 34.57 | 220.23 | 79.77 | 28.38 | 270.78 | 34.32 | 209.11 | 66.25 | 27.63 | 261.38 | 34.00 | 194.09 |
| 431 | 6 | 126.09 | 44.50 | 283.47 | 50.77 | 233.64 | 115.41 | 45.56 | 277.40 | 51.58 | 223.79 | 103.85 | 47.21 | 270.05 | 53.02 | 210.39 |
| 431 | 7 | 102.09 | 33.28 | 277.94 | 39.26 | 221.53 | 90.60 | 33.34 | 270.10 | 39.32 | 210.24 | 77.58 | 33.29 | 260.26 | 39.49 | 194.96 |
| 507 | 6 | 123.47 | 43.17 | 283.06 | 49.28 | 233.20 | 112.65 | 44.09 | 277.00 | 49.98 | 223.36 | 100.81 | 45.49 | 269.67 | 51.23 | 209.96 |
| 507 | 7 | 98.82 | 31.85 | 277.70 | 37.81 | 221.18 | 87.23 | 31.77 | 269.87 | 37.78 | 209.89 | 74.03 | 31.49 | 260.02 | 37.79 | 194.61 |
| 512 | 6 | 121.36 | 42.11 | 282.98 | 48.09 | 233.00 | 110.42 | 42.92 | 276.90 | 48.70 | 223.14 | 98.39 | 44.13 | 269.54 | 49.81 | 209.72 |
| 512 | 7 | 96.13 | 30.71 | 277.37 | 36.66 | 220.78 | 84.47 | 30.52 | 269.52 | 36.56 | 209.49 | 71.17 | 30.07 | 259.68 | 36.47 | 194.21 |
| 520 | 0-3 | 132.11 | 47.47 | 286.17 | 54.15 | 237.38 | 121.48 | 48.73 | 280.14 | 55.13 | 227.60 | 110.10 | 50.71 | 272.90 | 56.87 | 214.28 |
| 521 | 0-3 | 130.99 | 45.45 | 305.37 | 52.48 | 245.79 | 119.46 | 46.52 | 297.92 | 53.30 | 234.12 | 106.98 | 48.17 | 288.73 | 54.74 | 218.19 |
| 525 | 0-3 | 143.41 | 52.87 | 295.68 | 60.63 | 243.73 | 133.05 | 54.71 | 288.97 | 62.09 | 233.17 | 122.46 | 57.72 | 280.78 | 64.66 | 218.84 |
| 536 | 0-3 | 123.02 | 41.57 | 305.11 | 48.10 | 245.02 | 111.10 | 42.22 | 297.62 | 48.61 | 233.31 | 97.92 | 43.18 | 288.38 | 49.53 | 217.31 |
| 541 | 0-3 | 124.25 | 42.20 | 304.45 | 48.78 | 244.57 | 112.44 | 42.93 | 297.00 | 49.34 | 232.89 | 99.43 | 44.02 | 287.77 | 50.36 | 216.95 |
| 546 | 0-3 | 129.42 | 45.49 | 293.43 | 51.98 | 241.03 | 118.32 | 46.56 | 286.70 | 52.81 | 230.47 | 106.32 | 48.24 | 278.47 | 54.30 | 216.09 |
| 548 | 0-3 | 128.64 | 45.10 | 293.29 | 51.55 | 240.89 | 117.50 | 46.13 | 286.56 | 52.35 | 230.32 | 105.42 | 47.73 | 278.34 | 53.77 | 215.95 |
| 551 | 0-3 | 132.00 | 46.81 | 293.76 | 53.49 | 241.46 | 121.03 | 48.02 | 287.03 | 54.43 | 230.89 | 109.28 | 49.93 | 278.82 | 56.10 | 216.53 |
| 578 | 6 | 111.98 | 37.68 | 280.60 | 43.43 | 232.32 | 100.71 | 38.04 | 274.92 | 43.79 | 222.93 | 88.00 | 38.52 | 268.11 | 44.46 | 210.18 |
| 578 | 7 | 82.10 | 25.16 | 271.33 | 31.10 | 214.78 | 70.58 | 24.56 | 264.02 | 30.75 | 204.00 | 57.37 | 23.51 | 254.94 | 30.27 | 189.46 |
| 615 | 6 | 138.07 | 50.85 | 285.20 | 58.22 | 235.84 | 128.03 | 52.56 | 279.18 | 59.57 | 226.03 | 117.67 | 55.32 | 271.91 | 61.93 | 212.67 |
| 615 | 7 | 116.21 | 39.75 | 279.96 | 46.31 | 223.96 | 105.20 | 40.45 | 272.14 | 46.83 | 212.67 | 93.06 | 41.45 | 262.33 | 47.74 | 197.39 |
| 624 | 6 | 116.81 | 39.06 | 297.37 | 44.88 | 240.88 | 105.08 | 39.54 | 290.34 | 45.25 | 229.83 | 91.98 | 40.21 | 281.71 | 45.93 | 214.78 |
| 624 | 7 | 87.43 | 27.10 | 277.62 | 33.00 | 219.46 | 75.62 | 26.58 | 270.07 | 32.68 | 208.35 | 62.04 | 25.62 | 260.66 | 32.23 | 193.35 |
| 789 | 6 | 129.27 | 46.13 | 284.01 | 52.65 | 234.29 | 118.74 | 47.35 | 277.95 | 53.59 | 224.44 | 107.46 | 49.27 | 270.63 | 55.25 | 211.05 |
| 789 | 7 | 105.85 | 34.95 | 278.46 | 41.02 | 222.16 | 94.46 | 35.16 | 270.63 | 41.19 | 210.87 | 81.63 | 35.37 | 260.80 | 41.53 | 195.60 |
| 796 | 0-3 | 133.90 | 47.68 | 295.42 | 54.40 | 243.25 | 122.99 | 48.99 | 288.67 | 55.42 | 232.68 | 111.37 | 51.07 | 280.43 | 57.23 | 218.33 |
| 805 | 6 | 117.02 | 40.01 | 282.21 | 45.78 | 232.16 | 105.89 | 40.60 | 276.14 | 46.23 | 222.31 | 93.50 | 41.46 | 268.79 | 47.07 | 208.90 |
| 805 | 7 | 90.74 | 28.47 | 276.74 | 34.49 | 220.07 | 78.98 | 28.07 | 268.91 | 34.27 | 208.78 | 65.52 | 27.31 | 259.07 | 33.97 | 193.50 |
| 809 | 6 | 116.67 | 39.86 | 281.89 | 45.60 | 231.99 | 105.55 | 40.44 | 275.85 | 46.05 | 222.18 | 93.16 | 41.29 | 268.54 | 46.88 | 208.81 |
| 809 | 7 | 90.35 | 28.32 | 276.68 | 34.34 | 219.96 | 78.59 | 27.91 | 268.83 | 34.11 | 208.67 | 65.12 | 27.13 | 258.97 | 33.81 | 193.39 |
| 830 | 6 | 122.06 | 42.47 | 282.89 | 48.49 | 232.98 | 111.17 | 43.32 | 276.83 | 49.14 | 223.13 | 99.21 | 44.60 | 269.48 | 50.29 | 209.72 |
| 830 | 7 | 97.07 | 31.11 | 277.45 | 37.06 | 220.90 | 85.44 | 30.95 | 269.60 | 36.98 | 209.61 | 72.17 | 30.57 | 259.76 | 36.93 | 194.33 |
| 838 | 6 | 123.68 | 43.28 | 283.07 | 49.38 | 233.17 | 112.89 | 44.22 | 277.00 | 50.10 | 223.32 | 101.12 | 45.67 | 269.65 | 51.38 | 209.91 |
| 838 | 7 | 99.20 | 32.03 | 277.53 | 37.97 | 221.05 | 87.64 | 31.97 | 269.70 | 37.95 | 209.76 | 74.48 | 31.73 | 259.86 | 38.00 | 194.49 |
| 839 | 6 | 117.99 | 39.61 | 297.46 | 45.48 | 241.01 | 106.30 | 40.15 | 290.44 | 45.88 | 229.97 | 93.30 | 40.91 | 281.80 | 46.63 | 214.93 |
| 839 | 7 | 88.83 | 27.67 | 277.57 | 33.52 | 219.47 | 77.05 | 27.20 | 270.02 | 33.22 | 208.36 | 63.51 | 26.33 | 260.62 | 32.82 | 193.36 |
| 841 | 6 | 122.77 | 42.82 | 283.02 | 48.89 | 233.13 | 111.90 | 43.70 | 276.96 | 49.56 | 223.28 | 100.00 | 45.03 | 269.63 | 50.76 | 209.88 |
| 841 | 7 | 97.94 | 31.47 | 277.64 | 37.44 | 221.10 | 86.31 | 31.35 | 269.80 | 37.38 | 209.80 | 73.07 | 31.02 | 259.96 | 37.35 | 194.52 |
| 861 | 6 | 122.35 | 42.62 | 282.94 | 48.64 | 233.01 | 111.48 | 43.48 | 276.86 | 49.30 | 223.17 | 99.56 | 44.79 | 269.52 | 50.48 | 209.75 |
| 861 | 7 | 97.58 | 31.32 | 277.61 | 37.27 | 221.03 | 85.95 | 31.18 | 269.75 | 37.20 | 209.73 | 72.71 | 30.83 | 259.90 | 37.17 | 194.43 |
| 868 | 0-3 | 129.14 | 45.24 | 294.82 | 51.60 | 242.45 | 117.99 | 46.31 | 288.05 | 52.43 | 231.86 | 105.94 | 47.97 | 279.78 | 53.91 | 217.49 |
| 869 | 6 | 123.35 | 43.10 | 283.15 | 49.20 | 233.24 | 112.52 | 44.02 | 277.07 | 49.89 | 223.39 | 100.68 | 45.42 | 269.73 | 51.14 | 209.97 |
| 869 | 7 | 98.68 | 31.80 | 277.56 | 37.75 | 221.06 | 87.09 | 31.71 | 269.73 | 37.72 | 209.77 | 73.90 | 31.43 | 259.88 | 37.73 | 194.50 |
| 898 | 6 | 124.18 | 43.53 | 283.17 | 49.68 | 233.32 | 113.40 | 44.48 | 277.12 | 50.41 | 223.47 | 101.61 | 45.96 | 269.77 | 51.71 | 210.07 |
| 898 | 7 | 99.72 | 32.24 | 277.69 | 38.21 | 221.22 | 88.15 | 32.20 | 269.85 | 38.20 | 209.94 | 75.00 | 31.98 | 260.02 | 38.25 | 194.66 |
| 902 | 6 | 126.18 | 44.55 | 283.35 | 50.83 | 233.56 | 115.52 | 45.62 | 277.29 | 51.65 | 223.73 | 103.97 | 47.28 | 269.97 | 53.09 | 210.33 |
| 902 | 7 | 102.24 | 33.35 | 277.91 | 39.33 | 221.51 | 90.76 | 33.41 | 270.09 | 39.39 | 210.22 | 77.75 | 33.38 | 260.24 | 39.57 | 194.94 |
| 907 | 0-3 | 129.14 | 45.26 | 294.64 | 51.65 | 242.36 | 117.98 | 46.31 | 287.88 | 52.47 | 231.79 | 105.91 | 47.95 | 279.64 | 53.93 | 217.43 |
| 913 | 0-3 | 128.30 | 44.83 | 294.66 | 51.13 | 242.27 | 117.11 | 45.84 | 287.89 | 51.93 | 231.68 | 104.98 | 47.43 | 279.62 | 53.34 | 217.31 |
| 914 | 0-3 | 131.07 | 46.22 | 295.06 | 52.73 | 242.78 | 120.01 | 47.38 | 288.29 | 53.63 | 232.19 | 108.12 | 49.20 | 280.04 | 55.23 | 217.84 |
| 915 | 0-3 | 134.48 | 47.98 | 295.49 | 54.76 | 243.35 | 123.59 | 49.32 | 288.73 | 55.80 | 232.78 | 112.03 | 51.45 | 280.50 | 57.65 | 218.43 |
| 918 | 0-3 | 133.61 | 47.53 | 295.41 | 54.23 | 243.23 | 122.68 | 48.82 | 288.65 | 55.24 | 232.66 | 111.00 | 50.87 | 280.40 | 57.02 | 218.30 |
| 926 | 0-3 | 135.66 | 48.59 | 295.78 | 55.47 | 243.67 | 124.82 | 49.98 | 289.03 | 56.56 | 233.10 | 113.35 | 52.21 | 280.80 | 58.49 | 218.75 |
| 927 | 0-3 | 128.37 | 44.87 | 294.60 | 51.19 | 242.27 | 117.17 | 45.88 | 287.84 | 51.99 | 231.69 | 105.03 | 47.46 | 279.60 | 53.40 | 217.34 |
| 928 | 6 | 120.17 | 40.63 | 297.64 | 46.61 | 241.33 | 108.56 | 41.27 | 290.63 | 47.08 | 230.30 | 95.69 | 42.19 | 282.01 | 47.95 | 215.27 |
| 928 | 7 | 91.70 | 28.82 | 278.26 | 34.63 | 220.17 | 79.93 | 28.45 | 270.71 | 34.38 | 209.06 | 66.42 | 27.71 | 261.32 | 34.07 | 194.04 |
| 930 | 6 | 117.98 | 40.47 | 282.37 | 46.27 | 232.32 | 106.90 | 41.12 | 276.30 | 46.77 | 222.46 | 94.60 | 42.06 | 268.92 | 47.67 | 209.04 |
| 930 | 7 | 91.97 | 28.98 | 276.86 | 34.97 | 220.17 | 80.24 | 28.63 | 269.01 | 34.77 | 208.88 | 66.81 | 27.94 | 259.15 | 34.53 | 193.61 |
| 1479 | 0-3 | 126.22 | 43.13 | 304.73 | 49.84 | 244.93 | 114.49 | 43.96 | 297.27 | 50.47 | 233.25 | 101.63 | 45.22 | 288.04 | 51.61 | 217.31 |
| 1588 | 6 | 119.70 | 41.32 | 282.40 | 47.18 | 232.41 | 108.73 | 42.06 | 276.33 | 47.75 | 222.55 | 96.62 | 43.17 | 268.96 | 48.77 | 209.14 |

Table 6 - continued

| obsid | ccdid | gamma=1.2 | | | | | gamma=1.4 | | | | | gamma=1.7 | | | | |
|---|---|---|---|---|---|---|---|---|---|---|---|---|---|---|---|---|
| | | B | S | H | Sc | Hc | B | S | H | Sc | Hc | B | S | H | Sc | Hc |
| 1588 | 7 | 94.28 | 29.95 | 276.81 | 35.90 | 220.24 | 82.61 | 29.69 | 268.97 | 35.77 | 208.95 | 69.26 | 29.14 | 259.13 | 35.62 | 193.69 |
| 1596 | 6 | 121.51 | 42.21 | 282.66 | 48.17 | 232.72 | 110.63 | 43.04 | 276.59 | 48.81 | 222.87 | 98.67 | 44.31 | 269.24 | 49.95 | 209.46 |
| 1596 | 7 | 96.64 | 30.93 | 277.31 | 36.87 | 220.74 | 85.01 | 30.76 | 269.46 | 36.79 | 209.46 | 71.73 | 30.36 | 259.62 | 36.73 | 194.19 |
| 1629 | 6 | 123.96 | 43.43 | 282.99 | 49.54 | 233.13 | 113.19 | 44.38 | 276.93 | 50.28 | 223.29 | 101.45 | 45.86 | 269.60 | 51.58 | 209.89 |
| 1629 | 7 | 99.58 | 32.19 | 277.57 | 38.13 | 221.09 | 88.03 | 32.15 | 269.73 | 38.13 | 209.80 | 74.89 | 31.93 | 259.88 | 38.19 | 194.53 |
| 1630 | 6 | 124.53 | 43.72 | 283.05 | 49.87 | 233.22 | 113.79 | 44.70 | 276.99 | 50.63 | 223.38 | 102.10 | 46.23 | 269.65 | 51.97 | 209.98 |
| 1630 | 7 | 100.26 | 32.49 | 277.59 | 38.44 | 221.14 | 88.74 | 32.47 | 269.75 | 38.45 | 209.86 | 75.64 | 32.31 | 259.92 | 38.54 | 194.59 |
| 1642 | 6 | 121.34 | 42.12 | 282.65 | 48.08 | 232.70 | 110.45 | 42.95 | 276.58 | 48.71 | 222.85 | 98.47 | 44.20 | 269.22 | 49.84 | 209.44 |
| 1642 | 7 | 96.32 | 30.80 | 277.20 | 36.73 | 220.62 | 84.69 | 30.62 | 269.35 | 36.65 | 209.33 | 71.42 | 30.20 | 259.51 | 36.57 | 194.05 |
| 1643 | 6 | 124.09 | 43.50 | 282.99 | 49.62 | 233.15 | 113.33 | 44.46 | 276.93 | 50.36 | 223.31 | 101.60 | 45.94 | 269.60 | 51.67 | 209.91 |
| 1643 | 7 | 99.74 | 32.26 | 277.59 | 38.21 | 221.12 | 88.20 | 32.22 | 269.75 | 38.21 | 209.84 | 75.06 | 32.02 | 259.92 | 38.27 | 194.57 |
| 1644 | 6 | 121.88 | 42.39 | 282.73 | 48.37 | 232.80 | 111.01 | 43.24 | 276.67 | 49.02 | 222.95 | 99.09 | 44.54 | 269.33 | 50.19 | 209.54 |
| 1644 | 7 | 97.09 | 31.12 | 277.34 | 37.06 | 220.79 | 85.47 | 30.97 | 269.51 | 36.99 | 209.50 | 72.22 | 30.59 | 259.66 | 36.95 | 194.23 |
| 1708 | 0-3 | 128.35 | 44.85 | 294.66 | 51.18 | 242.32 | 117.15 | 45.87 | 287.89 | 51.97 | 231.73 | 105.00 | 47.44 | 279.64 | 53.38 | 217.36 |
| 1712 | 6 | 118.50 | 40.88 | 279.91 | 46.70 | 230.94 | 107.57 | 41.58 | 274.03 | 47.24 | 221.31 | 95.46 | 42.61 | 266.93 | 48.21 | 208.19 |
| 1712 | 7 | 92.05 | 29.15 | 272.98 | 35.07 | 217.12 | 80.50 | 28.84 | 265.26 | 34.90 | 205.98 | 67.27 | 28.23 | 255.56 | 34.70 | 190.91 |
| 2126 | 6 | 119.61 | 41.27 | 282.47 | 47.12 | 232.45 | 108.63 | 42.01 | 276.39 | 47.69 | 222.59 | 96.51 | 43.11 | 269.03 | 48.71 | 209.17 |
| 2126 | 7 | 94.13 | 29.89 | 276.79 | 35.84 | 220.21 | 82.46 | 29.62 | 268.95 | 35.70 | 208.93 | 69.11 | 29.07 | 259.11 | 35.55 | 193.67 |
| 2130 | 6 | 124.77 | 43.83 | 283.18 | 50.01 | 233.32 | 114.04 | 44.83 | 277.12 | 50.77 | 223.48 | 102.37 | 46.38 | 269.77 | 52.12 | 210.07 |
| 2130 | 7 | 100.55 | 32.61 | 277.69 | 38.56 | 221.23 | 89.03 | 32.61 | 269.85 | 38.58 | 209.94 | 75.94 | 32.46 | 260.00 | 38.69 | 194.67 |
| 2227 | 0-3 | 131.12 | 46.36 | 293.65 | 52.94 | 241.24 | 120.15 | 47.54 | 286.90 | 53.86 | 230.66 | 108.37 | 49.41 | 278.68 | 55.49 | 216.28 |
| 2228 | 0-3 | 129.55 | 45.56 | 293.27 | 52.03 | 240.85 | 118.50 | 46.67 | 286.54 | 52.89 | 230.27 | 106.59 | 48.40 | 278.31 | 54.42 | 215.90 |
| 2242 | 6 | 121.09 | 42.01 | 282.55 | 47.94 | 232.60 | 110.20 | 42.82 | 276.49 | 48.56 | 222.75 | 98.22 | 44.06 | 269.13 | 49.68 | 209.35 |
| 2242 | 7 | 96.04 | 30.68 | 277.06 | 36.62 | 220.49 | 84.41 | 30.50 | 269.22 | 36.53 | 209.20 | 71.13 | 30.06 | 259.37 | 36.45 | 193.93 |
| 2244 | 6 | 120.45 | 41.68 | 282.66 | 47.58 | 232.64 | 109.51 | 42.45 | 276.58 | 48.17 | 222.78 | 97.46 | 43.63 | 269.22 | 49.24 | 209.35 |
| 2244 | 7 | 95.25 | 30.35 | 277.05 | 36.29 | 220.47 | 83.59 | 30.12 | 269.22 | 36.18 | 209.18 | 70.27 | 29.63 | 259.37 | 36.07 | 193.91 |
| 2245 | 6 | 120.09 | 41.51 | 282.43 | 47.39 | 232.45 | 109.14 | 42.27 | 276.36 | 47.97 | 222.60 | 97.06 | 43.42 | 269.01 | 49.02 | 209.19 |
| 2245 | 7 | 94.75 | 30.14 | 276.88 | 36.09 | 220.31 | 83.10 | 29.91 | 269.05 | 35.97 | 209.03 | 69.77 | 29.39 | 259.21 | 35.84 | 193.76 |
| 2247 | 6 | 120.10 | 41.51 | 282.52 | 47.39 | 232.51 | 109.15 | 42.27 | 276.45 | 47.98 | 222.66 | 97.07 | 43.42 | 269.09 | 49.03 | 209.25 |
| 2247 | 7 | 94.74 | 30.14 | 276.90 | 36.08 | 220.32 | 83.08 | 29.90 | 269.06 | 35.96 | 209.03 | 69.75 | 29.38 | 259.23 | 35.83 | 193.77 |
| 2248 | 6 | 120.41 | 41.66 | 282.53 | 47.56 | 232.54 | 109.47 | 42.44 | 276.45 | 48.16 | 222.69 | 97.43 | 43.62 | 269.09 | 49.23 | 209.27 |
| 2248 | 7 | 95.16 | 30.31 | 277.00 | 36.25 | 220.41 | 83.51 | 30.09 | 269.16 | 36.14 | 209.13 | 70.19 | 29.59 | 259.31 | 36.03 | 193.85 |

Note. ECFs (10^-13 erg/sec/cm2 per 1 cnt/sec) are calculated in the following energy bands.
  B  = counts in 0.3-8.0 keV and flux in 0.3-8.0 keV
  S  = counts in 0.3-2.5 keV and flux in 0.5-2.0 keV
  H  = counts in 2.5-8.0 keV and flux in 2.0-8.0 keV
  Sc = counts in 0.5-2.0 keV and flux in 0.5-2.0 keV
  Hc = counts in 2.0-8.0 keV and flux in 2.0-8.0 keV

```
                                       Table 7 ChaMP X-ray Sources
--------------------------------------------------------------------------------------------
     source name         obsid ccdid src no    RA        DEC     error   D_off   radius eff_exp  flag
                                             (deg)      (deg)  (arcsec) (arcmin) (arcsec) (ksec)
--------------------------------------------------------------------------------------------

   CXOMP J001758.9+163119    520    2    20   4.495466  16.522075   2.2    8.2    17.1   59.01
   CXOMP J001801.7+163426    520    2    17   4.507208  16.573914   1.8    8.6    18.9   58.85
   CXOMP J001807.2+163551    520    2    15   4.530175  16.597551   2.3    8.4    18.0   58.78
   CXOMP J001807.9+163120    520    2     5   4.533049  16.522306   1.1    6.1     9.2   61.40
   CXOMP J001808.5+163231    520    2     4   4.535688  16.542151   1.2    6.3     9.7   61.61
   CXOMP J001809.3+162532    520    3    10   4.538769  16.425648   1.4    7.2    12.7   54.50
   CXOMP J001810.2+163223    520    2     3   4.542675  16.539993   1.1    5.9     8.6   60.89   032
   CXOMP J001810.2+162942    520    2    10   4.542813  16.495064   1.4    5.4     7.4   63.29
   CXOMP J001817.6+163107    520    2     2   4.573497  16.518677   1.0    3.8     4.2   65.27
   CXOMP J001818.0+163316    520    2     8   4.575296  16.554504   1.1    4.7     5.9   62.39

   CXOMP J001821.7+161941    520    3     9   4.590758  16.328226   2.9   10.7    28.8   51.51
   CXOMP J001825.0+163653    520    0    23   4.604235  16.614733   1.6    7.1    12.0   58.69
   CXOMP J001827.0+162900    520    3     4   4.612845  16.483385   1.0    1.7     3.0   60.45
   CXOMP J001828.5+162800    520    3     3   4.618802  16.466688   1.0    2.3     3.0   60.15
   CXOMP J001828.6+163418    520    0     8   4.619523  16.571732   1.0    4.3     5.2   58.49
   CXOMP J001831.4+162042    520    3     5   4.630925  16.345135   1.7    9.4    22.2   53.28
   CXOMP J001833.4+163154    520    0     3   4.639497  16.531778   1.0    1.8     3.0   64.21
   CXOMP J001836.8+163615    520    0    16   4.653358  16.604240   1.6    6.3     9.7   60.07   032
   CXOMP J001837.3+163447    520    0     2   4.655755  16.579763   1.0    4.9     6.1   58.53
   CXOMP J001837.4+163046    520    1     2   4.656161  16.512894   1.0    1.3     3.0   67.02

   CXOMP J001837.4+163757    520    0     7   4.656184  16.632698   1.6    8.0    16.1   56.87
   CXOMP J001837.5+163610    520    0    14   4.656549  16.603050   1.6    6.2     9.6   60.28   032
   CXOMP J001837.9+163910    520    0    13   4.658127  16.652940   2.0    9.2    21.4   55.63
   CXOMP J001838.1+163320    520    0     6   4.659020  16.555639   1.0    3.5     3.8   63.20
   CXOMP J001845.3+163528    520    0    10   4.689052  16.591274   1.8    6.2     9.5   59.57
   CXOMP J001845.7+163346    520    0     1   4.690596  16.562969   1.0    4.9     6.1   61.92
   CXOMP J001850.1+162756    520    1     1   4.708950  16.465687   1.0    4.7     5.8   62.97
   CXOMP J001853.5+162751    520    1     6   4.723330  16.464350   1.1    5.5     7.6   63.61
   CXOMP J001854.9+162952    520    1     5   4.728840  16.498003   1.1    5.3     7.3   62.55
   CXOMP J001859.8+162649    520    1     4   4.749342  16.447060   1.0    7.3    12.9   61.14

   CXOMP J001905.9+162842    520    1    12   4.774853  16.478407   2.1    8.1    16.4   59.25
   CXOMP J001909.2+163101    520    1    11   4.788618  16.517030   2.5    8.8    19.6   59.84
   CXOMP J005716.6-273230   2244    7    11  14.319263 -27.541681   1.2    5.0     6.5    6.45
   CXOMP J005717.9-271830   2242    7     3  14.324648 -27.308399   1.0    3.9     4.5    6.55
   CXOMP J005724.5-273201   2244    7     4  14.352160 -27.533709   1.0    3.5     3.8    6.83   052
   CXOMP J005724.5-273201   2242    6    18  14.352332 -27.534007   3.2    9.8    24.0    9.53   052
   CXOMP J005729.2-273043   2244    7     7  14.371962 -27.512104   1.0    3.5     3.8    6.73
   CXOMP J005730.8-273203   2244    7     3  14.378702 -27.534199   1.0    2.3     3.0    6.03   052
   CXOMP J005730.8-273203   2242    6     5  14.378766 -27.533735   3.5   10.1    25.4    9.10   052
   CXOMP J005732.8-273006   2244    7     6  14.386735 -27.501905   1.0    3.6     4.0    6.61

   CXOMP J005745.0-272922   2242    6     8  14.437648 -27.489685   2.8    9.2    21.6    9.53
   CXOMP J005759.9-272126   2245    7     1  14.499895 -27.357483   1.0    2.7     3.0    6.29
   CXOMP J005800.6-272741   2247    7    19  14.502738 -27.461517   1.5    6.1     9.2    9.72
   CXOMP J005803.4-272135   2245    7     2  14.514359 -27.359936   1.0    3.2     3.3    6.23
   CXOMP J005811.4-272635   2247    7    11  14.547684 -27.443153   1.0    4.6     5.8   10.06   055
   CXOMP J005813.9-272549   2247    7     9  14.558322 -27.430428   1.2    4.9     6.2   10.03
   CXOMP J005814.6-275002   2248    7    15  14.560858 -27.834078   1.4    5.3     7.2    9.53   055
   CXOMP J005819.9-272855   2247    7     2  14.583080 -27.482115   1.0    1.7     3.0   10.52   055
   CXOMP J005827.9-275157   2248    7     3  14.616590 -27.865887   1.0    3.5     3.8    8.94
   CXOMP J005828.0-275125   2248    7     9  14.616721 -27.857054   1.0    3.1     3.1    9.53

   CXOMP J005828.4-273033   2247    7     1  14.618617 -27.509281   1.0    0.8     3.0   10.53
   CXOMP J005834.9-272713   2247    7     5  14.645507 -27.453680   1.0    3.2     3.4   10.49
   CXOMP J005836.1-275016   2248    7     1  14.650720 -27.838053   1.0    1.1     3.0    9.63
   CXOMP J010117.1+315050    521    0     7  15.321324  31.847488   2.0   10.3    26.5   39.35
   CXOMP J010117.5+315157    521    0    14  15.322988  31.865944   2.8   10.5    27.9   37.87
   CXOMP J010123.9+314607    521    0     6  15.349731  31.768690   1.0    8.6    19.0   38.84   055
   CXOMP J010136.5+314655    521    0     5  15.402399  31.782181   1.1    5.8     8.6   42.51   032
   CXOMP J010136.9+315327    521    0    11  15.403825  31.891090   2.0    7.8    15.4   41.41
   CXOMP J010141.0+314503    521    2     8  15.420858  31.751110   1.0    5.6     8.0   42.81
   CXOMP J010148.4+314653    521    0     1  15.451965  31.781589   1.0    3.4     3.6   40.11

   CXOMP J010148.5+315348    521    1    23  15.452453  31.896822   1.7    6.6    10.5   42.77
   CXOMP J010151.7+314407    521    2     7  15.465645  31.735369   1.1    4.6     5.8   43.25
   CXOMP J010200.9+315224    521    1    10  15.503939  31.873590   1.0    4.4     5.4   45.28
   CXOMP J010204.0+315325    521    1     8  15.516984  31.890284   1.1    5.4     7.4   42.83
   CXOMP J010204.1+313921    521    2     5  15.517450  31.655953   1.8    8.7    19.2   39.59
   CXOMP J010207.0+314050    521    2     3  15.529260  31.680677   1.4    7.2    12.8   40.80
   CXOMP J010208.3+315638    521    1     5  15.534861  31.943981   1.6    8.7    19.1   41.07
   CXOMP J010214.1+314201    521    2     1  15.559063  31.700552   1.2    6.4    10.0   41.50
   CXOMP J010220.4+315110    521    1     2  15.585252  31.853022   1.0    4.8     6.0   44.97
   CXOMP J010222.6+315305    521    1    13  15.594188  31.884857   1.9    6.5    10.2   43.71
```

Table 8  X-ray Photometry

| source name | net(B) 0.3-8.0 | net(S1) 0.3-0.9 | net(S2) 0.9-2.5 | net(H) 2.5-8.0 | net(Sc) 0.5-2.0 | net(Hc) 2.0-8.0 | Flux(B) 0.3-8.0 |
|---|---|---|---|---|---|---|---|
| CXOMP J001758.9+163119 | 63.65 ( 11.39) | 4.54 ( 4.76) | 41.29 ( 8.30) | 17.83 ( 7.47) | 40.63 ( 8.38) | 20.20 ( 7.86) | 0.119 ( 0.021) |
| CXOMP J001801.7+163426 | 93.68 ( 12.94) | 4.22 ( 4.63) | 46.82 ( 8.92) | 42.65 ( 9.26) | 43.76 ( 8.72) | 49.15 ( 9.92) | 0.176 ( 0.024) |
| CXOMP J001807.2+163551 | 53.85 ( 10.83) | 1.73 ( 4.06) | 27.45 ( 7.40) | 24.68 ( 7.91) | 29.55 ( 7.55) | 26.10 ( 8.29) | 0.101 ( 0.020) |
| CXOMP J001807.9+163120 | 102.93 ( 11.79) | 7.21 ( 4.30) | 61.78 ( 9.12) | 33.94 ( 7.41) | 55.48 ( 8.81) | 47.22 ( 8.42) | 0.185 ( 0.021) |
| CXOMP J001808.5+163231 | 85.67 ( 11.08) | 11.88 ( 4.99) | 45.21 ( 8.07) | 28.58 ( 7.09) | 51.07 ( 8.55) | 32.54 ( 7.50) | 0.154 ( 0.020) |
| CXOMP J001809.3+162532 | 86.41 ( 11.66) | 12.52 ( 5.38) | 53.05 ( 8.76) | 20.84 ( 6.98) | 60.76 ( 9.46) | 24.43 ( 7.39) | 0.175 ( 0.024) |
| CXOMP J001810.2+163223 | 261.75 ( 17.98) | 61.37 ( 8.93) | 155.46 ( 13.88) | 44.92 ( 8.81) | 189.43 ( 15.07) | 61.61 ( 10.06) | 0.475 ( 0.033) |
| CXOMP J001810.2+162942 | 23.85 ( 6.94) | 7.24 ( 4.14) | 13.72 ( 5.23) | 2.90 ( 4.00) | 20.42 ( 5.99) | 3.77 ( 4.32) | 0.042 ( 0.012) |
| CXOMP J001817.6+163107 | 33.51 ( 7.07) | 0.47 ( 2.33) | 19.66 ( 5.56) | 13.37 ( 4.98) | 16.33 ( 5.22) | 17.23 ( 5.45) | 0.057 ( 0.012) |
| CXOMP J001818.0+163316 | 43.38 ( 8.00) | 5.16 ( 3.61) | 29.16 ( 6.55) | 9.06 ( 4.60) | 30.92 ( 6.73) | 12.70 ( 5.12) | 0.077 ( 0.014) |
| CXOMP J001821.7+161941 | 137.03 ( 17.23) | 29.30 ( 8.26) | 90.54 ( 12.31) | 17.19 ( 10.18) | 113.72 ( 13.55) | 19.98 ( 10.74) | 0.294 ( 0.037) |
| CXOMP J001825.0+163653 | 61.19 ( 9.98) | 5.43 ( 4.15) | 40.20 ( 7.78) | 15.56 ( 6.12) | 42.51 ( 8.00) | 19.39 ( 6.59) | 0.115 ( 0.019) |
| CXOMP J001827.0+162900 | 57.90 ( 8.80) | 5.72 ( 3.60) | 39.09 ( 7.39) | 13.09 ( 4.85) | 39.09 ( 7.39) | 18.81 ( 5.56) | 0.106 ( 0.016) |
| CXOMP J001828.5+162800 | 144.04 ( 13.16) | 23.55 ( 5.98) | 84.79 ( 10.32) | 35.69 ( 7.15) | 100.44 ( 11.14) | 42.59 ( 7.70) | 0.265 ( 0.024) |
| CXOMP J001828.6+163418 | 52.59 ( 8.55) | 10.42 ( 4.43) | 33.22 ( 6.90) | 8.96 ( 4.44) | 40.83 ( 7.54) | 10.96 ( 4.72) | 0.099 ( 0.016) |
| CXOMP J001831.4+162042 | 213.44 ( 17.88) | 25.25 ( 7.29) | 148.56 ( 14.08) | 39.63 ( 9.71) | 159.13 ( 14.57) | 54.50 ( 10.80) | 0.443 ( 0.037) |
| CXOMP J001833.4+163154 | 220.07 ( 15.93) | 46.56 ( 7.91) | 135.47 ( 12.70) | 38.03 ( 7.31) | 160.23 ( 13.72) | 54.94 ( 8.54) | 0.379 ( 0.027) |
| CXOMP J001836.8+163615 | 36.14 ( 7.51) | 1.03 ( 2.97) | 23.69 ( 5.89) | 11.43 ( 5.01) | 21.14 ( 5.69) | 15.59 ( 5.49) | 0.067 ( 0.014) |
| CXOMP J001837.3+163447 | 125.47 ( 12.49) | 36.32 ( 7.15) | 70.74 ( 9.53) | 18.41 ( 5.79) | 97.55 ( 10.99) | 25.22 ( 6.47) | 0.237 ( 0.024) |
| CXOMP J001837.4+163046 | 36.03 ( 7.23) | 7.70 ( 3.96) | 22.41 ( 5.88) | 5.92 ( 3.79) | 24.36 ( 6.08) | 11.67 ( 4.72) | 0.059 ( 0.012) |
| CXOMP J001837.4+163757 | 133.05 ( 14.07) | 17.29 ( 6.01) | 96.22 ( 11.36) | 19.54 ( 7.31) | 95.48 ( 11.41) | 36.59 ( 8.74) | 0.259 ( 0.027) |
| CXOMP J001837.5+163610 | 43.72 ( 8.17) | 0.00 ( 2.97) | 30.91 ( 6.65) | 12.94 ( 5.13) | 28.70 ( 6.56) | 15.79 ( 5.49) | 0.080 ( 0.015) |
| CXOMP J001837.9+163910 | 166.76 ( 16.07) | 24.33 ( 7.03) | 99.42 ( 11.80) | 43.00 ( 9.69) | 118.15 ( 12.70) | 46.21 ( 10.15) | 0.332 ( 0.032) |
| CXOMP J001838.1+163320 | 47.36 ( 8.13) | 6.71 ( 3.79) | 32.28 ( 6.81) | 8.37 ( 4.29) | 36.47 ( 7.15) | 10.98 ( 4.72) | 0.083 ( 0.014) |
| CXOMP J001845.3+163528 | 28.34 ( 7.59) | 1.20 ( 3.21) | 5.58 ( 4.15) | 21.55 ( 6.49) | 3.68 ( 3.99) | 25.32 ( 6.93) | 0.053 ( 0.014) |
| CXOMP J001845.7+163346 | 294.13 ( 18.41) | 24.65 ( 6.18) | 180.49 ( 14.56) | 88.99 ( 10.69) | 187.25 ( 14.82) | 105.22 ( 11.54) | 0.525 ( 0.033) |
| CXOMP J001850.1+162756 | 111.06 ( 11.87) | 23.07 ( 6.09) | 67.51 ( 9.36) | 20.47 ( 5.89) | 80.37 ( 10.16) | 29.28 ( 6.74) | 0.195 ( 0.021) |
| CXOMP J001853.5+162751 | 174.33 ( 14.61) | 8.61 ( 4.45) | 107.31 ( 11.53) | 58.41 ( 9.00) | 91.06 ( 10.80) | 80.92 ( 10.33) | 0.303 ( 0.025) |
| CXOMP J001854.9+162952 | 61.92 ( 9.38) | 16.62 ( 5.34) | 35.24 ( 7.15) | 10.06 ( 4.87) | 43.92 ( 7.85) | 16.48 ( 5.69) | 0.110 ( 0.017) |
| CXOMP J001859.8+162649 | 367.21 ( 20.80) | 53.11 ( 8.68) | 247.15 ( 16.94) | 66.95 ( 9.96) | 253.84 ( 17.19) | 101.26 ( 11.79) | 0.664 ( 0.038) |
| CXOMP J001905.9+162842 | 66.43 ( 11.11) | 3.59 ( 4.32) | 37.37 ( 7.87) | 25.47 ( 7.75) | 32.49 ( 7.65) | 33.30 ( 8.46) | 0.124 ( 0.021) |
| CXOMP J001909.2+163101 | 62.77 ( 12.18) | 13.25 ( 6.16) | 33.77 ( 8.06) | 15.75 ( 8.07) | 41.53 ( 8.82) | 18.61 ( 8.57) | 0.116 ( 0.023) |
| CXOMP J005716.6-273230 | 31.98 ( 9.88) | 9.38 ( 5.00) | 14.73 ( 6.42) | 7.87 ( 6.91) | 17.63 ( 6.69) | 11.43 ( 7.48) | 0.350 ( 0.108) |
| CXOMP J005717.9-271830 | 36.57 ( 7.15) | 17.90 ( 5.33) | 13.90 ( 4.84) | 4.76 ( 3.40) | 20.86 ( 5.67) | 5.76 ( 3.60) | 0.399 ( 0.078) |
| CXOMP J005724.5-273201 | 80.55 ( 10.87) | 34.00 ( 7.07) | 37.37 ( 7.56) | 9.18 ( 5.38) | 46.76 ( 8.21) | 11.56 ( 5.82) | 0.833 ( 0.112) |
| CXOMP J005724.5-273201 | 50.32 ( 9.72) | 8.30 ( 4.77) | 33.31 ( 7.44) | 8.71 ( 5.68) | 35.71 ( 7.84) | 14.01 ( 6.39) | 0.519 ( 0.100) |
| CXOMP J005729.2-273043 | 30.50 ( 7.97) | 13.80 ( 5.11) | 13.22 ( 5.36) | 3.48 ( 4.76) | 26.55 ( 6.65) | 4.05 ( 5.03) | 0.320 ( 0.084) |
| CXOMP J005730.8-273203 | 65.92 ( 9.67) | 28.87 ( 6.55) | 21.65 ( 5.99) | 15.39 ( 5.58) | 38.65 ( 7.47) | 17.76 ( 5.90) | 0.772 ( 0.113) |
| CXOMP J005730.8-273203 | 52.46 ( 9.88) | 16.14 ( 5.60) | 31.81 ( 7.36) | 4.51 ( 5.34) | 44.46 ( 8.39) | 7.44 ( 5.90) | 0.567 ( 0.107) |
| CXOMP J005732.8-273006 | 72.34 ( 10.57) | 12.46 ( 4.99) | 44.01 ( 8.07) | 15.87 ( 6.22) | 55.67 ( 8.88) | 16.70 ( 6.41) | 0.773 ( 0.113) |
| CXOMP J005745.0-272922 | 65.10 ( 9.97) | 13.01 ( 5.11) | 37.04 ( 7.48) | 15.05 ( 5.81) | 40.13 ( 7.78) | 22.03 ( 6.58) | 0.671 ( 0.103) |
| CXOMP J005759.9-272126 | 172.63 ( 14.18) | 114.95 ( 11.76) | 54.95 ( 8.47) | 2.72 ( 2.94) | 138.91 ( 12.82) | 4.72 ( 3.40) | 1.923 ( 0.158) |
| CXOMP J005800.6-272741 | 35.58 ( 7.40) | 16.56 ( 5.34) | 14.97 ( 5.10) | 4.05 ( 3.62) | 26.66 ( 6.37) | 3.69 ( 3.62) | 0.257 ( 0.053) |
| CXOMP J005803.4-272135 | 55.69 ( 8.54) | 26.00 ( 6.17) | 22.00 ( 5.77) | 7.69 ( 3.96) | 39.00 ( 7.31) | 8.69 ( 4.13) | 0.626 ( 0.096) |
| CXOMP J005811.4-272635 | 55.38 ( 8.61) | 33.72 ( 6.90) | 18.63 ( 5.45) | 3.03 ( 3.19) | 42.44 ( 7.62) | 2.98 ( 3.19) | 0.386 ( 0.060) |
| CXOMP J005813.9-272549 | 44.93 ( 7.92) | 19.17 ( 5.56) | 17.86 ( 5.33) | 7.89 ( 4.13) | 27.59 ( 6.36) | 11.85 ( 4.72) | 0.314 ( 0.055) |
| CXOMP J005814.6-275002 | 27.06 ( 6.65) | 13.54 ( 4.98) | 13.97 ( 4.98) | 0.00 ( 2.35) | 25.65 ( 6.28) | 0.00 ( 2.35) | 0.200 ( 0.049) |
| CXOMP J005819.9-272855 | 28.53 ( 6.46) | 20.86 ( 5.67) | 7.00 ( 3.78) | 0.67 ( 2.33) | 21.91 ( 5.77) | 0.67 ( 2.33) | 0.190 ( 0.043) |
| CXOMP J005827.9-275157 | 68.16 ( 9.35) | 28.58 ( 6.46) | 33.81 ( 6.90) | 5.77 ( 3.60) | 56.62 ( 8.60) | 7.72 ( 3.96) | 0.538 ( 0.074) |
| CXOMP J005828.0-275125 | 28.66 ( 6.46) | 22.81 ( 5.88) | 6.00 ( 3.60) | 0.00 ( 1.87) | 23.95 ( 5.98) | 0.00 ( 1.87) | 0.212 ( 0.048) |
| CXOMP J005828.4-273033 | 26.48 ( 6.27) | 12.81 ( 4.71) | 9.81 ( 4.28) | 3.86 ( 3.18) | 17.86 ( 5.33) | 4.76 ( 3.40) | 0.176 ( 0.042) |
| CXOMP J005834.9-272713 | 30.39 ( 6.64) | 5.81 ( 3.60) | 20.81 ( 5.67) | 3.77 ( 3.18) | 22.67 ( 5.88) | 7.77 ( 3.96) | 0.203 ( 0.044) |
| CXOMP J005836.1-275016 | 36.32 ( 7.15) | 13.76 ( 4.84) | 18.85 ( 5.45) | 3.71 ( 3.18) | 25.81 ( 6.17) | 6.66 ( 3.79) | 0.266 ( 0.052) |
| CXOMP J010117.1+315050 | 294.83 ( 20.87) | 80.95 ( 10.99) | 159.67 ( 14.65) | 54.21 ( 11.49) | 212.05 ( 16.54) | 69.06 ( 12.51) | 0.805 ( 0.057) |
| CXOMP J010117.5+315157 | 92.79 ( 15.32) | 10.91 ( 6.68) | 51.78 ( 9.99) | 30.10 ( 10.66) | 57.53 ( 10.44) | 34.84 ( 11.34) | 0.263 ( 0.043) |
| CXOMP J010123.9+314607 | 967.85 ( 32.90) | 221.33 ( 16.22) | 583.33 ( 25.45) | 163.19 ( 14.78) | 720.22 ( 28.13) | 212.02 ( 16.60) | 2.677 ( 0.091) |
| CXOMP J010136.5+314655 | 143.79 ( 12.97) | 28.78 ( 6.28) | 74.59 ( 9.71) | 40.42 ( 7.41) | 94.17 ( 10.64) | 50.48 ( 8.08) | 0.363 ( 0.033) |
| CXOMP J010136.9+315327 | 68.36 ( 11.28) | 10.07 ( 5.26) | 43.40 ( 8.30) | 14.89 ( 6.98) | 45.32 ( 8.57) | 22.29 ( 7.78) | 0.177 ( 0.029) |
| CXOMP J010141.0+314503 | 310.08 ( 18.96) | 65.58 ( 9.30) | 189.46 ( 14.89) | 55.04 ( 8.88) | 226.05 ( 16.16) | 80.15 ( 10.39) | 0.778 ( 0.048) |
| CXOMP J010148.4+314653 | 179.21 ( 14.49) | 43.60 ( 7.69) | 99.42 ( 11.04) | 36.19 ( 7.15) | 133.25 ( 12.61) | 44.02 ( 7.77) | 0.480 ( 0.039) |
| CXOMP J010148.5+315348 | 33.56 ( 8.49) | 5.48 ( 4.16) | 21.18 ( 6.20) | 6.89 ( 5.53) | 24.14 ( 6.58) | 9.52 ( 5.96) | 0.084 ( 0.021) |
| CXOMP J010151.7+314407 | 34.02 ( 7.32) | 8.03 ( 4.13) | 17.60 ( 5.45) | 8.39 ( 4.45) | 25.36 ( 6.28) | 8.96 ( 4.59) | 0.085 ( 0.018) |
| CXOMP J010200.9+315224 | 75.46 ( 10.00) | 2.08 ( 2.95) | 45.74 ( 7.92) | 27.63 ( 6.56) | 33.65 ( 6.99) | 40.91 ( 7.70) | 0.179 ( 0.024) |
| CXOMP J010204.0+315325 | 87.20 ( 10.86) | 29.30 ( 6.64) | 52.42 ( 8.48) | 5.49 ( 4.31) | 71.87 ( 9.71) | 9.75 ( 5.00) | 0.219 ( 0.027) |
| CXOMP J010204.1+313921 | 114.55 ( 14.06) | 21.72 ( 6.88) | 55.04 ( 9.42) | 37.79 ( 9.17) | 69.63 ( 10.44) | 43.30 ( 9.71) | 0.311 ( 0.038) |
| CXOMP J010207.0+314050 | 162.35 ( 14.71) | 35.60 ( 7.48) | 99.20 ( 11.35) | 27.55 ( 7.35) | 123.60 ( 12.54) | 35.52 ( 8.10) | 0.427 ( 0.039) |
| CXOMP J010208.3+315638 | 203.74 ( 17.04) | 49.14 ( 8.83) | 115.25 ( 12.33) | 39.35 ( 9.32) | 136.91 ( 13.35) | 61.66 ( 10.76) | 0.533 ( 0.045) |
| CXOMP J010214.1+314201 | 101.17 ( 11.84) | 5.63 ( 3.98) | 72.03 ( 9.77) | 23.51 ( 6.76) | 71.09 ( 9.78) | 30.41 ( 7.43) | 0.262 ( 0.031) |
| CXOMP J010220.4+315110 | 61.24 ( 9.25) | 3.60 ( 3.41) | 40.65 ( 7.55) | 16.99 ( 5.57) | 37.18 ( 7.31) | 24.68 ( 6.38) | 0.146 ( 0.022) |
| CXOMP J010222.6+315305 | 29.48 ( 7.74) | 3.34 ( 3.62) | 18.61 ( 5.79) | 7.53 ( 5.14) | 17.98 ( 5.79) | 12.46 ( 5.82) | 0.072 ( 0.019) |

```
                         Table 9  X-ray colors
--------------------------------------------------------------------------------
      source name               HR              C21             C32
                             (0.3-8.0)       (0.3-2.5)       (0.9-8.0)
--------------------------------------------------------------------------------

CXOMP J001758.9+163119      -0.44 (  0.20)   -0.96 (  4.22)   0.36 (  0.09)
CXOMP J001801.7+163426      -0.09 (  0.14)   -1.05 (  5.38)   0.04 (  0.11)
CXOMP J001807.2+163551      -0.08 (  0.21)   -1.20 ( 16.36)   0.05 (  0.16)
CXOMP J001807.9+163120      -0.34 (  0.13)   -0.93 (  2.29)   0.26 (  0.06)
CXOMP J001808.5+163231      -0.33 (  0.14)   -0.58 (  0.75)   0.20 (  0.08)
CXOMP J001809.3+162532      -0.52 (  0.16)   -0.63 (  0.85)   0.41 (  0.06)
CXOMP J001810.2+163223      -0.61 (  0.08)   -0.44 (  0.20)   0.48 (  0.03)
CXOMP J001810.2+162942      -0.76 (  0.38)   -0.28 (  0.57)   0.68 (  0.13)
CXOMP J001817.6+163107      -0.20 (  0.23)  -99.00 (  0.00)   0.17 (  0.14)
CXOMP J001818.0+163316      -0.58 (  0.22)   -0.75 (  1.80)   0.51 (  0.07)

CXOMP J001821.7+161941      -0.75 (  0.16)   -0.49 (  0.42)   0.72 (  0.05)
CXOMP J001825.0+163653      -0.49 (  0.19)   -0.87 (  2.54)   0.41 (  0.07)
CXOMP J001827.0+162900      -0.55 (  0.18)   -0.83 (  1.95)   0.47 (  0.06)
CXOMP J001828.5+162800      -0.50 (  0.11)   -0.56 (  0.44)   0.38 (  0.04)
CXOMP J001828.6+163418      -0.66 (  0.20)   -0.50 (  0.66)   0.57 (  0.06)
CXOMP J001831.4+162042      -0.63 (  0.10)   -0.77 (  0.78)   0.57 (  0.03)
CXOMP J001833.4+163154      -0.65 (  0.09)   -0.46 (  0.25)   0.55 (  0.03)
CXOMP J001836.8+163615      -0.42 (  0.31)  -99.00 (  0.00)   0.39 (  0.12)
CXOMP J001837.3+163447      -0.71 (  0.12)   -0.29 (  0.20)   0.58 (  0.04)
CXOMP J001837.4+163046      -0.67 (  0.25)   -0.46 (  0.73)   0.58 (  0.08)

CXOMP J001837.4+163757      -0.71 (  0.13)   -0.75 (  0.89)   0.69 (  0.04)
CXOMP J001837.5+163610      -0.51 (  0.26)  -99.00 (  0.00)   0.49 (  0.09)
CXOMP J001837.9+163910      -0.48 (  0.11)   -0.61 (  0.55)   0.36 (  0.05)
CXOMP J001838.1+163320      -0.65 (  0.21)   -0.68 (  1.26)   0.59 (  0.06)
CXOMP J001845.3+163528       0.52 (  0.32)   -0.67 (  5.59)  -0.59 (  1.34)
CXOMP J001845.7+163346      -0.40 (  0.07)   -0.86 (  0.84)   0.31 (  0.03)
CXOMP J001850.1+162756      -0.63 (  0.13)   -0.47 (  0.38)   0.52 (  0.04)
CXOMP J001853.5+162751      -0.33 (  0.09)   -1.10 (  2.86)   0.26 (  0.04)
CXOMP J001854.9+162952      -0.68 (  0.19)   -0.33 (  0.35)   0.54 (  0.07)
CXOMP J001859.8+162649      -0.64 (  0.07)   -0.67 (  0.36)   0.57 (  0.02)

CXOMP J001905.9+162842      -0.23 (  0.18)   -1.02 (  5.53)   0.17 (  0.11)
CXOMP J001909.2+163101      -0.50 (  0.22)   -0.41 (  0.58)   0.33 (  0.12)
CXOMP J005716.6-273230      -0.51 (  0.35)   -0.20 (  0.30)   0.27 (  0.43)
CXOMP J005717.9-271830      -0.74 (  0.17)    0.11 (  0.20)   0.47 (  0.34)
CXOMP J005724.5-273201      -0.77 (  0.12)   -0.04 (  0.13)   0.61 (  0.27)
CXOMP J005724.5-273201      -0.65 (  0.20)   -0.60 (  0.27)   0.58 (  0.30)
CXOMP J005729.2-273043      -0.77 (  0.28)    0.02 (  0.24)   0.58 (  0.62)
CXOMP J005730.8-273203      -0.53 (  0.14)    0.13 (  0.16)   0.15 (  0.20)
CXOMP J005730.8-273203      -0.83 (  0.19)   -0.29 (  0.18)   0.85 (  0.52)
CXOMP J005732.8-273006      -0.56 (  0.14)   -0.55 (  0.19)   0.44 (  0.19)

CXOMP J005745.0-272922      -0.54 (  0.15)   -0.46 (  0.19)   0.39 (  0.19)
CXOMP J005759.9-272126      -0.97 (  0.03)    0.32 (  0.08)   1.31 (  0.47)
CXOMP J005800.6-272741      -0.77 (  0.19)    0.04 (  0.20)   0.57 (  0.41)
CXOMP J005803.4-272135      -0.72 (  0.13)    0.07 (  0.15)   0.46 (  0.25)
CXOMP J005811.4-272635      -0.89 (  0.11)    0.26 (  0.16)   0.79 (  0.47)
CXOMP J005813.9-272549      -0.65 (  0.16)    0.03 (  0.18)   0.35 (  0.26)
CXOMP J005814.6-275002      -1.00 (  0.18)   -0.01 (  0.22)  99.00 (  0.00)
CXOMP J005819.9-272855      -0.95 (  0.16)    0.47 (  0.26)  99.00 (  0.00)
CXOMP J005827.9-275157      -0.83 (  0.10)   -0.07 (  0.13)   0.77 (  0.28)
CXOMP J005828.0-275125      -1.00 (  0.13)    0.58 (  0.28)  99.00 (  0.00)

CXOMP J005828.4-273033      -0.71 (  0.22)    0.12 (  0.25)   0.41 (  0.41)
CXOMP J005834.9-272713      -0.75 (  0.19)   -0.55 (  0.29)   0.74 (  0.39)
CXOMP J005836.1-275016      -0.80 (  0.16)   -0.14 (  0.20)   0.71 (  0.39)
CXOMP J010117.1+315050      -0.63 (  0.09)   -0.29 (  0.14)   0.47 (  0.03)
CXOMP J010117.5+315157      -0.35 (  0.18)   -0.68 (  1.32)   0.24 (  0.10)
CXOMP J010123.9+314607      -0.66 (  0.04)   -0.42 (  0.10)   0.55 (  0.01)
CXOMP J010136.5+314655      -0.49 (  0.11)   -0.46 (  0.35)   0.34 (  0.05)
CXOMP J010136.9+315327      -0.56 (  0.20)   -0.63 (  1.04)   0.47 (  0.07)
CXOMP J010141.0+314503      -0.65 (  0.07)   -0.46 (  0.20)   0.54 (  0.02)
CXOMP J010148.4+314653      -0.60 (  0.10)   -0.36 (  0.21)   0.44 (  0.04)

CXOMP J010148.5+315348      -0.59 (  0.30)   -0.59 (  1.36)   0.49 (  0.12)
CXOMP J010151.7+314407      -0.51 (  0.25)   -0.34 (  0.57)   0.32 (  0.13)
CXOMP J010200.9+315224      -0.27 (  0.14)   -1.34 ( 13.62)   0.22 (  0.08)
CXOMP J010204.0+315325      -0.87 (  0.17)   -0.25 (  0.22)   0.98 (  0.04)
CXOMP J010204.1+313921      -0.34 (  0.13)   -0.40 (  0.40)   0.16 (  0.09)
CXOMP J010207.0+314050      -0.66 (  0.11)   -0.45 (  0.29)   0.56 (  0.04)
CXOMP J010208.3+315638      -0.61 (  0.10)   -0.37 (  0.21)   0.47 (  0.04)
CXOMP J010214.1+314201      -0.54 (  0.14)   -1.11 (  4.00)   0.49 (  0.04)
CXOMP J010220.4+315110      -0.45 (  0.17)   -1.05 (  4.74)   0.38 (  0.07)
CXOMP J010222.6+315305      -0.49 (  0.31)   -0.75 (  2.73)   0.39 (  0.13)
```